\documentclass{pazha2}
\usepackage{graphicx} 
\usepackage{multirow}
\usepackage{amsmath}
\usepackage{color}
\definecolor{darkblue}{rgb}{0,0,0.9}

\def\*{$^{*}$}
\def\aa{$^{\mbox{\footnotesize a}}$}
\def\bb{$^{\mbox{\footnotesize b}}$}
\def\cc{$^{\mbox{\footnotesize c}}$}
\def\dd{$^{\mbox{\footnotesize d}}$}
\def\ee{$^{\mbox{\footnotesize e}}$}
\def\ff{$^{\mbox{\footnotesize f}}$}
\def\gg{$^{\mbox{\footnotesize g}}$}
\def\hh{$^{\mbox{\footnotesize h}}$}

\begin{document}
\journalinfo{\!}{2020}{46}{4}{205}{231}{251}[223]

\title{\bf Morphology of the Light Curves for the X-ray Novae H\,1743-322
and GX\,339-4 during their Outbursts in 2005--2019}
\year=2020
\author{A.~S.~Grebenev\email{\small grebenev@iki.rssi.ru}\address{1},
Yu.~A.~Dvorkovich\address{1}, V.~S.~Knyazeva\address{1},
K.~D.~Ostashenko\address{1}, S.~A.~Grebenev\address{2},
I.~A.~Mereminskiy\address{2}, and A.~V.~Prosvetov\address{2}
\addrestextf{1}{Moscow School 444, Nizhnyaya Pervomaiskaya ul. 14, Moscow, Russia}
\addrestextl{2}{Space Research Institute, Russian Academy of Sciences, Profsoyuznaya ul. 84/32, Moscow, 117997 Russia}
}

\shortauthor{GREBENEV et al.}
\shorttitle{MORPHOLOGY OF THE LIGHT CURVES FOR THE X-RAY NOVAE} 

\submitted{December 2, 2019}
\revised{December 16, 2019}
\accepted{December 21, 2019}

\begin{abstract}
\noindent
Based on long-term SWIFT, RXTE, and MAXI observations of the
X-ray novae H\,1743-322 (IGR\,J17464-3213) and GX\,339-4, we
have investigated the morphology and classified the light curves
of their X-ray outbursts. In particular, we have confirmed the
existence of two radically different types of outbursts, soft
(S) and hard (H), in both sources and revealed their varieties,
ultrabright (U) and intermediate (I). The properties and origin
of the differences in the light curves of these outbursts are
discussed in terms of the truncated accretion disk model.\\

\noindent
{\bf DOI:} 10.1134/S1063773720040052\\

{\bf Keywords:\/} {black holes, low-mass X-ray binaries, X-ray
  transients, X-ray novae, nonstationary accretion.}
\end{abstract}

\sloppypar
\emergencystretch=5pt
\looseness=-2
\raggedbottom
\parskip=4.5pt

\section*{INTRODUCTION}
\noindent
Nonstationary (flaring) X-ray binaries in which a stellar-mass
($M_1 \la 10\ M_{\odot}$) black hole (or a neutron star with a
weak, $B< 10^9$ G, magnetic field) and a low-mass ($M_2 \la
M_{\odot}$) main-sequence star serve as a compact object and a
normal component, respectively, are called X-ray novae. During
their outbursts X-ray novae become the brightest sources in the
X-ray sky (Sunyaev et al. 1988, 1991; Grebenev et al. 1993,
1997; Tanaka and Shibazaki 1996; Grove et al. 1998; Remillard
and McClintock 2006; Belloni 2010).

As a rule, the orbital periods of X-ray novae are several hours
(Cherepashchuk 2013). This ensures that the Roche lobe is filled
(or almost filled) by the normal star and makes an efficient
transfer of its matter through the inner Lagrange point (L1)
possible. The energy release through the accretion of this
matter by the black hole (neutron star) feeds the binary's
X-ray outburst. Since the accreting matter has a large angular
momentum, accretion occurs with the formation of an extended
accretion disk around the compact object in which the matter
slowly spirals in toward the center.

At present, it is unclear whether the outbursts of X-ray novae
are associated with some processes in the normal star leading to
its swelling to the Roche lobe volume or the mass transfer
occurs continually, but most of the time the matter does not
reach the black hole, but is accumulated in the outer disk
regions and flows inward only on reaching some critical
mass. Typical X-ray nova outbursts last for months; the duration
of especially powerful ones can reach one year. During the
quiescent (off) state the flux from novae drops below the
detection level by wide-field X-ray telescopes and all-sky
monitors. Nevertheless, the observations of some known X-ray
novae by telescopes with mirror optics have shown that the X-ray
flux does not disappear completely, but only drops by 4--5
orders of magnitude. The intervals between outbursts in some
novae are tens of years or more. Previously unknown sources of
this type are discovered almost every year, i.e., at the epoch
of X-ray astronomy they flare up for the first time. In other
novae outbursts occur quasi-regularly on a time scale of one or
two years. However, we know only a few such recurrent transients,
whereas the total number of observed X-ray novae approaches
already fifty (see, e.g., Cherepashchuk 2013).

The matter that ``broke through'' to the black hole as a result
of the nonstationary accretion episode ``spreads'' over the
accretion disk to form the observed X-ray light curve
(Lyubarskii and Shakura 1987; Lipunova and Shakura 2000;
Suleimanov et al. 2008).  As a rule, for powerful outbursts in
the standard 2--10 keV X-ray band it has a characteristic shape
called FRED (fast rise--exponential decay). Hard (\mbox{$>20$}
keV) radiation is observed on the first days and on the descent
of the light curve. However, there are also differences both
between the outbursts of different sources and between the
individual outbursts of the same object. Outbursts during which
hard radiation dominated throughout its duration are known
(Sunyaev et al. 1991; Grove et al. 1998; Mereminskiy et
al. 2017).

Hardness variations reflect the fact that in the course of
outburst development the X-ray novae alternately pass several
different states differing by their X-ray energy spectra: hard,
intermediate hard, soft, intermediate soft, and two-component
ones (Makishima et al. 1986; Grebenev et al. 1997; Belloni
2010). Clearly, these states are associated with different
regimes of disk accretion being realized in these binary systems
at a given specific time.

The variety of spectral states of X-ray novae finds a natural
explanation in the so-called truncated disk model. According to
this model, the accretion disk consists of two geometrically
separated regions: outer --- cold ($kT\la 1$ keV), geometrically
thin, opaque and, hence, radiating like a blackbody (it is
responsible for the soft component in the X-ray spectrum of the
nova; see Shakura and Sunyaev 1973) and inner ---
high-temperature ($kT\la 100$ keV), geometrically thick, but
optically thin --- semitransparent (it is responsible for the
hard power-law X-ray emission component; see, e.g., Shapiro et
al. 1976; Sunyaev and Truemper 1979). The transition between the
regions is determined by the evaporation of the cold outer disk
under the action of viscous energy release in it. According to
the model, one or the other state of the X-ray nova is observed,
depending on the position of the boundary between the outer and
inner regions (and, accordingly, their surface areas and
contributions to the total radiation spectrum). It would seem
that the radius of the boundary between the regions should
increase with accretion rate (the area of the hot zone should
grow with energy release; see Shakura and Sunyaev
1973). However, observations show that this does not happen, at
least there is no direct proportionality here.

In this paper we make an attempt to check the validity of the
described picture of the state transition and the truncated disk
model in general by studying the X-ray light curves for the
outbursts of two recurrent, widely known X-ray novae,
H\,1743-322 and GX\,339-4, observed in them over almost 15
years.  Most of these outbursts were investigated to some extent
by various instruments directly during their development and
decay, but, as far as we know, nobody has compared the
properties of different outbursts and analyzed their entire set.

\section*{THE SOURCES H\,1743-322 AND GX\,339-4}
\noindent
\underline{\sl The X-ra\!}{\sl \,y}\!\underline{\sl
  \ \,transient H\,1743-322\/} was discovered by the
\mbox{HEAO-1} observatory during its powerful 1977 outburst that
lasted $\sim230$ days. However, its position in the sky was
determined ambiguously and is given in the catalog of the main
HEAO-1 A1 instrument (Wood et al. 1984) with a large offset (the
source in this catalog was even called differently,
H\,1741-322). After this outburst the source had remained in the
off state for many years. A new outburst from this region was
recorded only in March 2003 (Revnivtsev et al. 2003) by the
INTEGRAL observatory (Winkler et al. 2003) and was mistaken for
the outburst of a new transient named IGR\,J17464-3213 due to
the position error in the catalog by Wood et al. (1984). The
association with H\,1743-322 was clarified by Markwardt (2003),
who pointed to the mentioning of a different possible position
of H\,1741-322, which coincides within the error limits with the
position of IGR\,J17464-3213, in the paper by Gursky et
al. (1978) based on HEAO-1 A3 data.
\begin{figure*}[t]
\centerline{\includegraphics[width=0.99\linewidth]{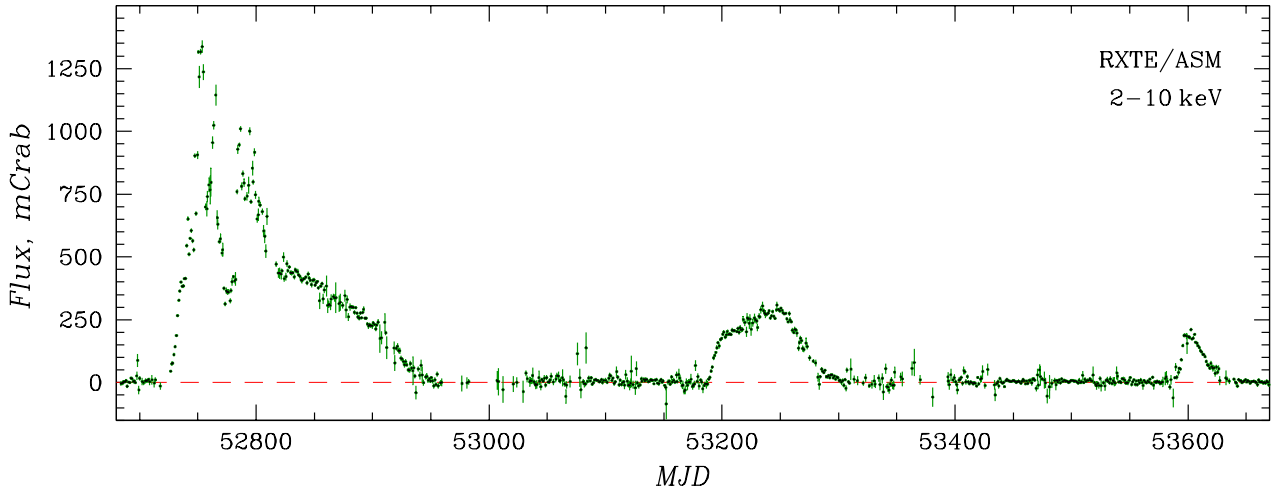}}

\caption{\rm Light curve for the giant 2003 outburst and the two
  succeeding outbursts of the X-ray nova H\,1743-322 in the
  2--10 keV energy band (according to the RXTE/ASM measurements
  from February 2003 to October 2005, MJD\,52680--53670).
\label{fig:lc-h1743.2003}}
\end{figure*}

The 2003 outburst was very powerful, lasted eight months, and
passed through the entire spectrum of states --- from hard
(Revnivtsev et al. 2003) to soft (Kretschmar et al. 2003) and
back (Grebenev et al. 2003; Tomsick and Kalemci 2003). Based on
RXTE data (Jahoda et al. 1996), Swank (2004) reported the onset
of a new outburst of the source in July 2004. A faint outburst
was also recorded by RXTE in August 2005 (Swank et al. 2005). It
was also observed by the Neil Gehrels SWIFT observatory (Gehrels et
al. 2004). The source's light curve measured by the all-sky
monitor (ASM) of the RXTE observatory during these three
outbursts is shown in Fig.\,\ref{fig:lc-h1743.2003}. The first
two outbursts differ greatly in power, duration, and time
profile from the third and all the succeeding outbursts observed
since that time every 8--9 months.  In this paper we will
investigate the third and succeeding outbursts, which are more
typical for this X-ray nova.

The optical counterpart of H\,1743-322 has not yet been
established. The assumption about accretion onto a black hole is
based on the commonality of the X-ray properties of this source
with the observational manifestations of other X-ray novae, in
particular on the observation of radio and X-ray jets from it
(Corbel et al. 2005). Since the source's position in the sky is
close to the Galactic center direction, it is generally believed
that it is actually located near the center at a distance
$d\simeq 8$ kpc. Steiner et al. (2012) estimated the distance to
the source more rigorously by analyzing the data on the jet
kinematics: $d=8.5\pm0.8$ kpc.

\underline{\sl The X-ra\!}{\sl \,y}\!\underline{\sl
  \ \,transient GX\,339-4\/} (also designated as
4U\,1658-48), like the X-ray source Cyg\,X-1, is a long-known
and best-studied (canonical) Galactic binary system containing a
black hole. In contrast to Cyg\,X-1, GX\,339-4 is a member of a
low-mass X-ray binary. The source is also the first known
recurrent transient whose outbursts are repeated on a time scale
of several years.

Discovered by the OSO-7 satellite in 1973 as an unusual highly
variable X-ray source (Markert et al. 1973), GX\,339-4 was
studied in detail by the ARIEL-6 (Motch et al. 1983), TENMA
(Makishima et al. 1986), GINGA (Miyamoto et al. 1991), GRANAT
(Grebenev et al. 1991, 1993) satellites and all of the
succeeding missions.
\begin{figure*}[t]
\centerline{\includegraphics[width=0.95\textwidth]{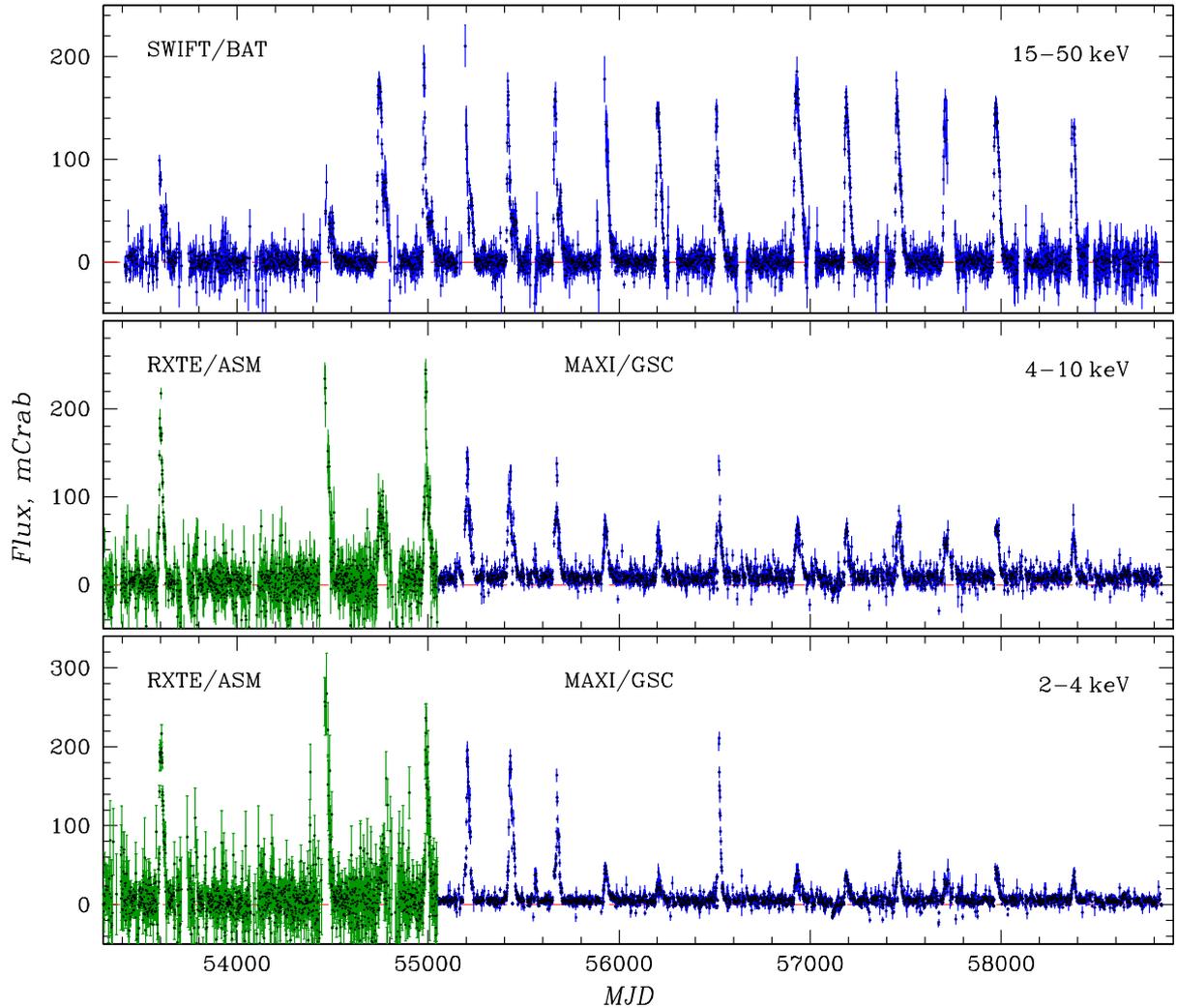}}

\caption{\rm Long-term light curve of the X-ray nova H\,1743-322
  in three energy bands (according to the SWIFT/BAT, MAXI/GSC
  (blue), and RXTE/ASM (green) measurements from January 2005 to
  December 2019, MJD 53300--58900). Seventeen nova outbursts of
  different types occurred over this period. Each point
  represents the data averaged over one day; as a rule, the real
  observations are noticeably shorter.
\label{fig:lc-h1743}}
\end{figure*}

The optical star V\,821~Ara of this binary system revolves
around the compact object with a 1.76-day period (Cherepashchuk
2013); the mass function of the compact object $f(M_1)=5.8\pm
0.5\ M_{\odot}$ (Hynes et al. 2003) leaves no doubt that it is a
black hole.  The distance to the source is estimated to lie
within the range $6\ \mbox{\rm kpc}<d<15$ kpc (Hynes et
al. 2004).\\ [-1mm]
\begin{table*}[t] 
\caption{Parameters of the outbursts of the X-ray nova
  H\,1743-322 from the SWIFT/BAT, RXTE/ASM,  and\protect\\ MAXI/GSC data\label{table:h1743}}

\vspace{3mm}
\small
\hspace{0cm}\begin{tabular}{c|r@{--}l|c|c|c|c|r@{}l|r@{}l|c|c}\hline
\multicolumn{3}{c|}{}& \multicolumn{2}{c|}{}&  &  & &  & & &\\ [-3mm]
\multicolumn{3}{c|}{Outburst\aa}&\multicolumn{2}{c|}{$T_{\rm m}$\bb}&\multicolumn{1}{c|}{$\Delta  T$\cc}&$T_{\rm r}$\dd&\multicolumn{2}{c|}{$F_{\rm m}(H)$\ee}&\multicolumn{2}{c|}{$F_{\rm m}(S)$\ee}&$SHR$\ff&Type\gg\\ \cline{1-11}
&\multicolumn{2}{c|}{}&  & & & & & & \multicolumn{2}{c|}{} &  \\ [-3mm]
N&\multicolumn{2}{c|}{MJD}&\multicolumn{1}{c|}{YYYY--MM}&MJD&days&days&\multicolumn{2}{c|}{mCrab}&\multicolumn{2}{c|}{mCrab}&&\\ \hline
&\multicolumn{2}{c|}{}& & & & && & & & &\\ [-3.5mm]
1& \multicolumn{2}{c|}{2}&\multicolumn{1}{c|}{3}&\multicolumn{1}{c}{4}& \multicolumn{1}{c}{5}&6 & \multicolumn{2}{c}{7}& \multicolumn{2}{c}{8}&9&10\\ \hline 
&\multicolumn{3}{c|}{}& & & &  & & & & \\ [-3.6mm]
1&53590&53640&2005--08&53595&50&--    & $98.95$&$\pm5.58$    &$216.71$&$\pm11.09$ &$2.19\pm0.17$ &U\\
2&54464&54512&2008--01&54469&48&874& $77.61$&$\pm16.50$  &$267.34$&$\pm51.15$ &$3.44\pm0.98$ &U\\
3&54732&54804&2008--10&54744&72&268& $177.69$&$\pm8.26$  &$160.00$&$\pm34.09$ &$0.90\pm0.20$ &I1\\
4&54973&55036&2009--05&54979&63&241& $193.57$&$\pm9.62$  &$236.46$&$\pm17.55$ &$1.22\pm0.11$ &S\\
5&55195&55245&2010--01&55195&50&222& $210.33$&$\pm20.43$&$194.66$&$\pm6.96$   &$0.93\pm0.10$ &S\\
6&55411&55471&2010--08&55418&60&216& $176.44$&$\pm8.28$  &$188.48$&$\pm8.98$   &$1.07\pm0.07$ &S\\
~7\hh&55535&55577&2011--01&55565&42&124&$<42$&$\ (3\sigma)$&$39.38$&$\pm8.46$&$>0.94\ (3\sigma)$&M\\
8&55656&55712&2011--04&55667&56&121  & $165.74$&$\pm9.88$ &$164.04$&$\pm8.60$   &$0.99\pm0.08$&I2\\
9&55924&55972&2011--12&55924&48&268& $178.13$&$\pm22.47$&$48.99$&$\pm5.13$     &$0.28\pm0.05$ &H\\
10&56191&56233&2012--10&56199&42&267& $149.75$&$\pm6.86$  &$38.46$&$\pm9.06$   &$0.26\pm0.06$ &H\\
11&56500&56556&2013--08&56510&56&309& $151.18$&$\pm7.89$  &$210.96$&$\pm8.52$ &$1.39\pm0.09$ &S\\
12&56911&56974&2014--10&56930&63&411& $185.51$&$\pm14.80$&$43.97$&$\pm8.40$   &$0.24\pm0.05$ &H\\
13&57177&57229&2015--06&57188&52&266& $165.32$&$\pm6.75$  &$41.28$&$\pm4.48$   &$0.25\pm0.03$ &H\\
14&57442&57500&2016--03&57452&58&265& $176.78$&$\pm9.05$  &$64.93$&$\pm4.77$   &$0.37\pm0.03$ &H\\
15&57694&57760&2016--11&57707&66&252& $158.66$&$\pm10.30$&$43.09$&$\pm8.75$   &$0.27\pm0.06$ &H\\
16&57957&58013&2017--08&57971&56&263& $154.64$&$\pm7.38$  &$47.72$&$\pm5.00$   &$0.31\pm0.04$ &H\\
17&58361&58404&2018--09&58383&43&404&$131.65$&$\pm8.44$   &$42.20$&$\pm5.17$   &$0.32\pm0.04$ &H\\
\hline
\multicolumn{13}{l}{}\\ [-1mm]
\multicolumn{13}{l}{\aa\ \footnotesize The outburst number and start-end dates (MJD).}\\
\multicolumn{13}{l}{\bb\ \footnotesize The year, month, and date (MJD) of
  reaching the peak flux in the outburst.}\\
\multicolumn{13}{l}{\cc\ \footnotesize The outburst duration.}\\
\multicolumn{13}{l}{\dd\ \footnotesize The recurrence time (the interval between the onset of this outburst and the onset of the previous one).}\\ 
\multicolumn{13}{l}{\ee\ \footnotesize The peak flux in the hard (H: 15--50
  keV) and soft (S: 2--4 or 1.3--3 keV during the MAXI/GSC}\\
\multicolumn{13}{l}{\ \ \  \footnotesize and RXTE/ASM observations, respectively) bands.}\\ 
\multicolumn{13}{l}{\ff\ \footnotesize The radiation softness (the ratio of the fluxes in the S and H bands).}\\
\multicolumn{13}{l}{\gg\ \footnotesize The type of outburst (hard -- H, soft -- S,
  ultrasoft -- U, intermediate -- I, micro-outburst -- M).}\\
\multicolumn{13}{l}{\hh\ \footnotesize The timing characteristics for this outburst were determined from the soft (S) band.}\\
\end{tabular}
\end{table*}

\section*{INSTRUMENTS AND OBSERVATIONS}
\noindent
The work is based on the archival publicly accessible data of
observations of these X-ray sources by the Neil Gehrels SWIFT (Gehrels et
al. 2004) (data from the BAT hard X-ray telescope with a coded
aperture were used; Barthelmy et al. 2005), RXTE (Jahoda et
al. 1996) (data from the ASM X-ray all-sky monitor were used),
and MAXI (Matsuoka et al. 2009) (onboard the International Space
Station (ISS), data from the GSC X-ray all-sky monitor were used;
Mihara et al. 2011) orbital astrophysical observatories. These
instruments have a wide field of view and carry out
quasi-regular observations of many bright cosmic sources.

The light curves of the novae H\,1743-322 and \mbox{GX\,339-4}
in the hard 15--50 keV band were taken from the site {\sl
  $<\,$swift.gsfc.nasa.gov/results/transients$\,>$\/}. They were
constructed from the SWIFT/BAT data by the telescope designers
within the framework of an X-ray transient monitoring program
(Krimm et al. 2013). Each point of the light curves is the
day-averaged flux based on at least 64~s of observations of 
the source. The light curves span the time interval from
February 12, 2005, to December 24, 2019.  A complete detailed
description of the light curve construction procedure and the
limitations taken into account in it can be found in the
mentioned paper by Krimm et al. (2013).

The light curves in the softer X-ray bands before MJD\,55200
were taken from the site \mbox{\sl $<\,$xte.mit.edu/asm\underline{~}lc.html$\,>$\/} of the
RXTE/ASM monitor.  They were provided by the instrument
designers in the 1.3--3 and 3--12.2 keV bands. After MJD\,55200
(from August 15, 2009, to December 24, 2019) we used data from
the MAXI/GSC monitor: the light curves in the 2--4 and 4--10 keV
bands were taken from the site {\sl
  $<\,$maxi.riken.jp/top/slist.html$\,>$\/}. Each point on the light
curves of both instruments is the day-averaged flux from the
source. It is based on 5--10 scans with a duration $\sim90$~s for
RXTE/ASM and several scans with a duration $\sim60$~s performed
every 92 min for MAXI/GSC.

For all instruments the measured photon fluxes were converted to
mCrabs ($10^{-3}$ of the flux from the Crab nebula) based on the
results of regular observations of the nebula by these
instruments presented at the above sites.\\ [-3mm]
\begin{figure}[t]
\centerline{\includegraphics[width=0.96\linewidth]{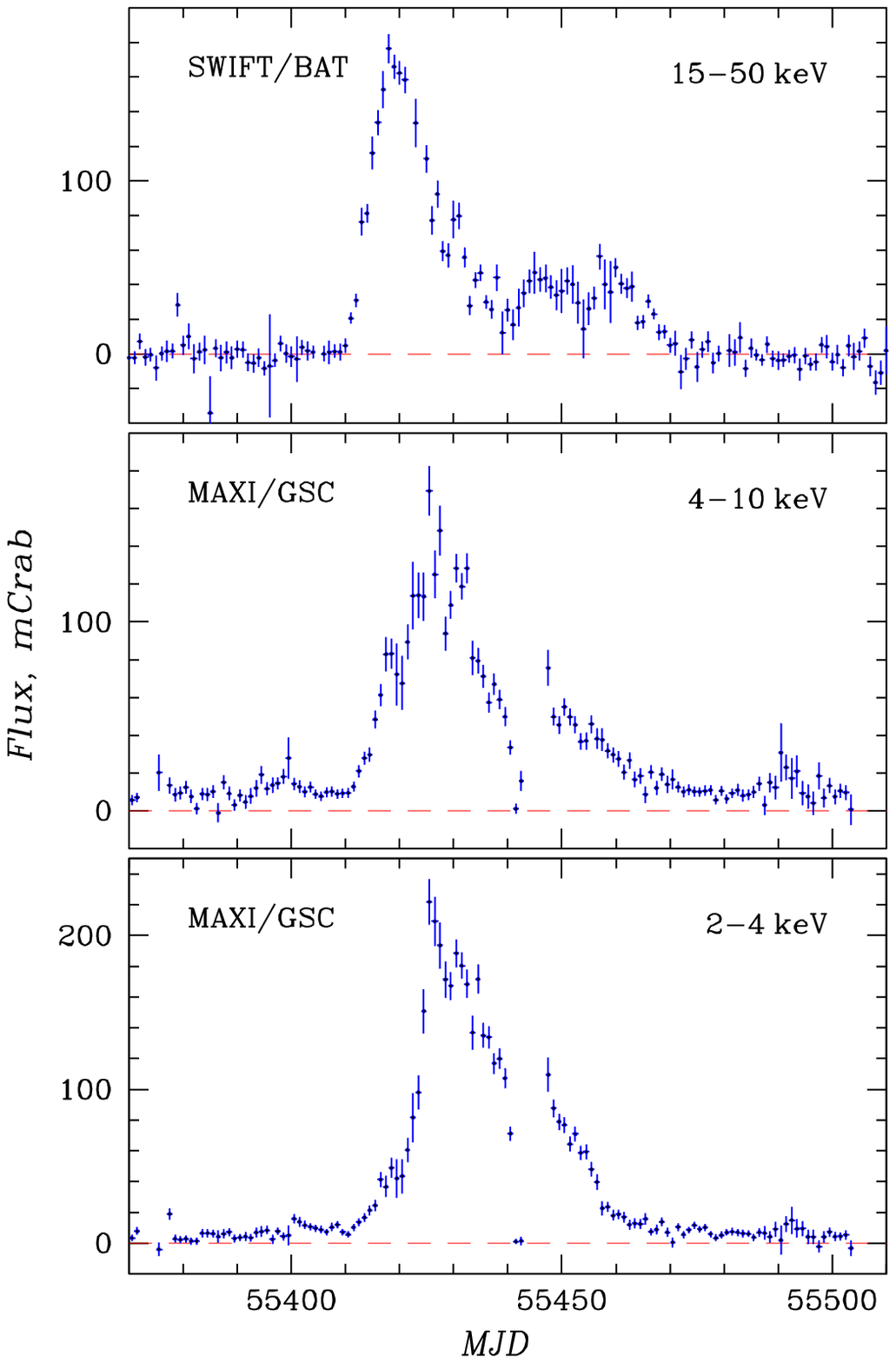}}

\caption{\rm Light curve for the outburst of the X-ray nova
  H\,1743-322 in August 2010 (outburst 6 in Table\,1, type S) in
  various energy bands (according to the SWIFT/BAT and MAXI/GSC
  measurements).\label{fig:lc-h1743.2010}}
\end{figure}
\begin{figure}[t]
\centerline{\includegraphics[width=0.96\linewidth]{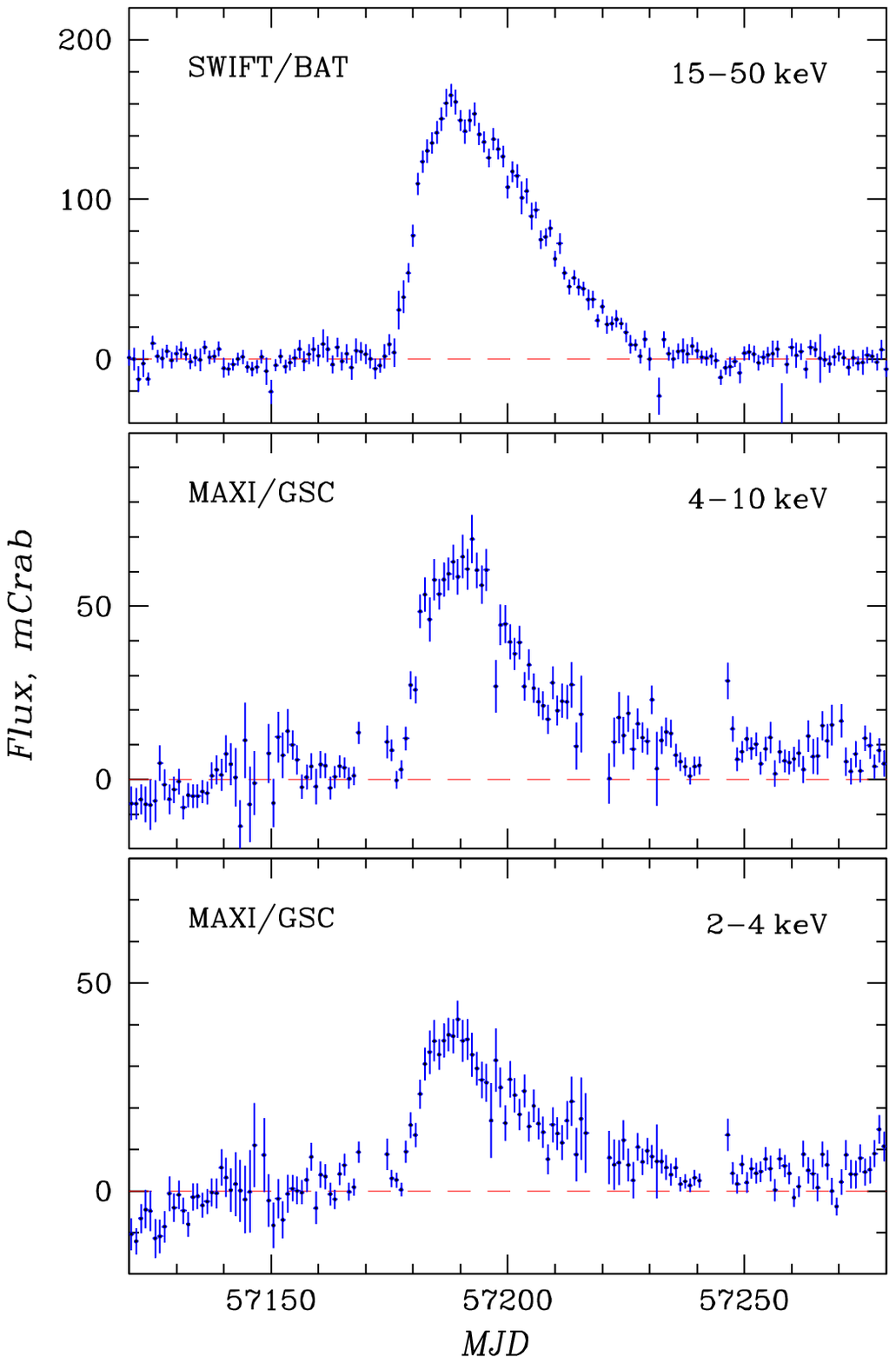}}

\caption{\rm Same as Fig.\,\protect\ref{fig:lc-h1743.2010}, but
  for the 2015 outburst (outburst 13, type H). The outburst
  amplitude in the 2--4 and 4--10 keV bands decreased
  noticeably.\label{fig:lc-h1743.2015}}
\end{figure}

\section*{RESULTS}
\noindent
Figure\,\ref{fig:lc-h1743} shows long-term (2005--2019) light
curves of the X-ray nova H\,1743-322 in three energy bands, 2--4
and 4--10 keV (from ISS/MAXI/GSC or RXTE/ASM data) and 15--50
keV (from SWIFT/BAT data).

In $\sim15$ years elapsed since January 2005 the source
experienced 17 outbursts. The last outburst occurred in 2018
(Grebenev and Mereminskiy 2018; Williams et al. 2020). It
follows from the figure that in the hard band, except for the
first five years of its observations by SWIFT (during which two
moderately strong, $\la50$ mCrab, outbursts occurred), all of the
outbursts had virtually the same intensity, reaching
$\sim150-200$ mCrab at maximum. The situation in soft bands,
4--10 keV and especially 2--4 keV, is different; here, in
addition to three ultrabright (with a flux $\sim 250-300$ mCrab)
and four moderately bright (with a flux $\sim 200$
mCrab) outbursts, outbursts whose peak flux did not exceed
$\sim50$ mCrab were regularly observed. Clearly, the radiation
from the source during these outbursts was much harder than that
during the bright ones.

Table\,1 gives some parameters of the outbursts from this
source: the outburst number and start-end date (MJD) (columns 1
and 2, respectively), the year and month of reaching the peak
flux (3), MJD of the peak flux (4), the outburst duration in
days (5), the recurrence time (the time between the onset of a
given outburst and the onset of the previous one) (6), the peak
flux in the hard (H: 15--50 keV) and soft (S: 2--4 keV) bands
(columns 7 and 8), and the softness $SHR$ of the radiation from
the source (the ratio of the fluxes in the soft and hard bands,
column 9). The timing characteristics were determined from the
light curve in the hard band (except for outburst 7). In the
soft band they slightly differ, in particular, the peak flux, as
a rule, is reached several days later than in the hard
one. However, the accuracy of these characteristics does not
exceed a few days due to the statistical errors of measurement
of the light curves and is insufficient for reliable conclusions
about their energy dependence to be drawn. Note that many
outbursts exhibit a quasi-recurrence with a period $\sim 260$
days.\\ [-1mm]

Obviously, the outbursts of this X-ray nova can be divided at
least into two most commonly encountered types --- soft (S) and
hard (H). Figures\,\ref{fig:lc-h1743.2010} and
\ref{fig:lc-h1743.2015} present typical profiles for outbursts
of these types (outbursts 6 and 13 from Table\,1). $SHR$ (column
9) lies in the range 0.9--1.4 for the soft outbursts and
0.24--0.37 for the hard ones. Apart from a several times greater
amplitude of the peak fluxes in the soft 2--4 and 4--10 keV
bands, the S outbursts are characterized by a more complex shape
of the light curve in the hard 15--50 keV band: after reaching
its peak, the flux decreases quite rapidly (on a time scale of
20--30 days) by a factor of 4--5 and then is kept at this level
for another $\sim30$ days. During an H outburst the photon flux
in the hard band decays noticeably more slowly (on a time scale
of 50--60 days), while the outburst light curve itself has a
simple shape without features resembling FRED. For each outburst
column 10 in the table gives its type.

\begin{figure}[t]
\centerline{\includegraphics[width=0.96\linewidth]{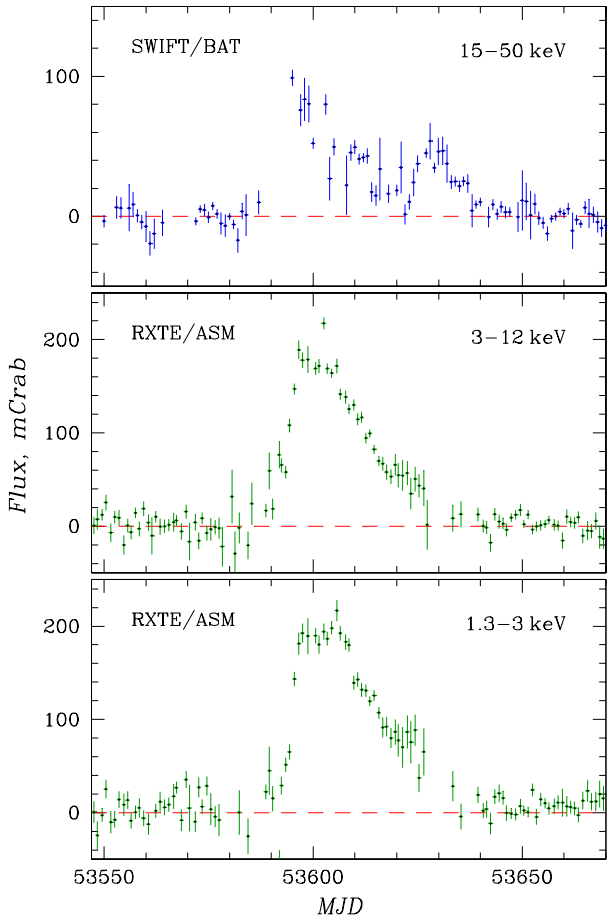}}

\caption{\rm Same as Fig.\,\protect\ref{fig:lc-h1743.2010}, but
  for the outburst of the X-ray nova H\,1743-322 in 2005
  (outburst 1, type U). This outburst should have formally been
  assigned to the S type by noting that its amplitude in the
  soft bands reaches a maximum. However, in the hard 15--50 keV
  band the outburst is much fainter than for other S outbursts, not
  to mention the H outbursts. Note that in the soft bands the
  measurements were performed by RXTE/ASM.\label{fig:lc-h1743.2005}}
\end{figure}

The two early (before 2008) outbursts are designated in the
table as U (ultrasoft) ones. In many respects they are similar
to the S outbursts, but are slightly brighter in the soft bands
than the remaining outbursts of this type and much (by a factor
of 2--3) fainter in the hard one. Their softness $SHR$ falls
within the range 2.2--3.5. Figure\,\ref{fig:lc-h1743.2005}
presents a light curve for the 2005 outburst of this type
(outburst 1 from Table\,1). The 2009 outburst 4 (assigned to the
S type) is as bright in the soft bands as the U outbursts, but
in the hard band it is comparable to the remaining S outbursts
(the softness of its radiation $1.22\pm0.11$ is near the upper
boundary of the range of values for the S outbursts).

If the entire activity period of the X-ray nova H\,1743-322,
from 2005 up until now, is described, then it can be seen that
before 2011 the source experienced mostly soft U and S
outbursts, while beginning since 2011 it completely passed to H
outbursts.

\begin{figure}[th]

\centerline{\includegraphics[width=0.96\linewidth]{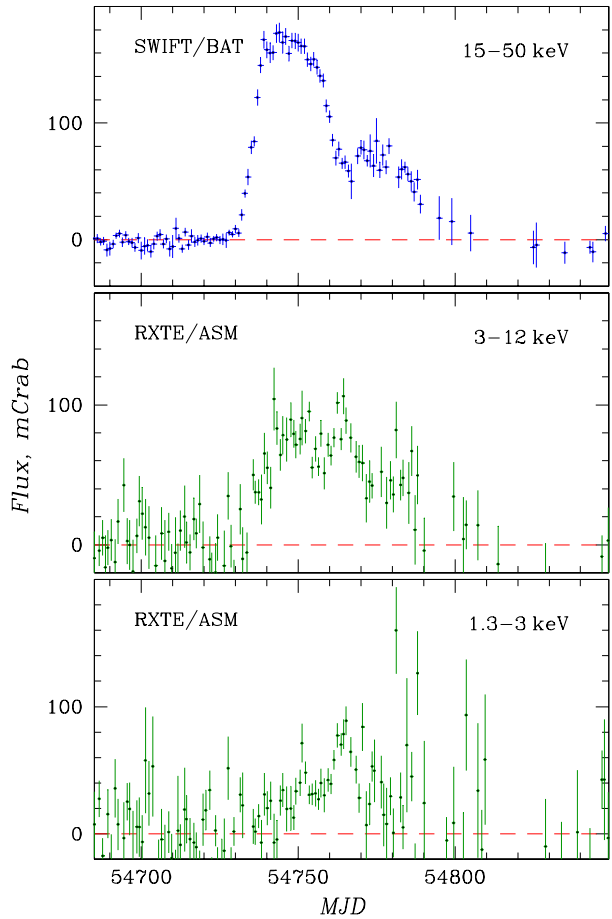}}

\caption{\rm Same as Fig.\,\protect\ref{fig:lc-h1743.2010}, but
  for the outburst in October 2008 (outburst 3, type I). The
  outburst should have formally be assigned to the S type, but
  in the soft bands, especially in 1.3--3 keV, a flux during the
  outburst lower than that during other outbursts of this type
  by a factor of 1.5--2 was recorded.\label{fig:lc-h1743.2008.10}}
\end{figure}

\begin{figure}[t]

\centerline{\includegraphics[width=0.96\linewidth]{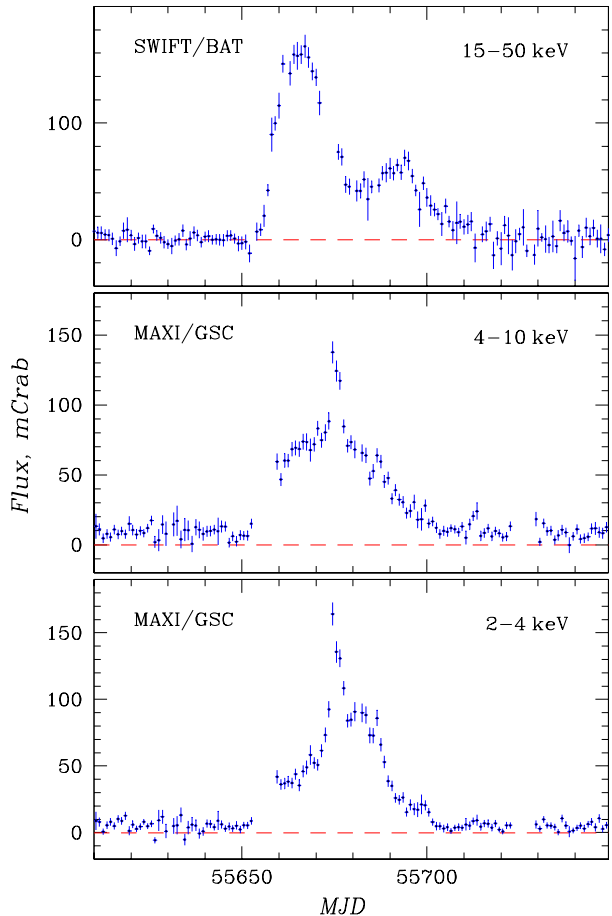}}

\caption{\rm Same as Fig.\,\protect\ref{fig:lc-h1743.2010}, but
  for the outburst in April 2011 (outburst 8, type I). The
  outburst should have formally be assigned to the S type, but
  in the soft bands, were it not for the short (3--4 days)
  radiation pulse, it would be much closer to the H
  outbursts.\label{fig:lc-h1743.2011.04}}
\end{figure}

It can be seen from Table\,1 that there are also other more rare
types of outbursts (in the table they are designated as I1, I2,
and M). Types I (intermediate) are shown in
Figs.\,\ref{fig:lc-h1743.2008.10} and
\ref{fig:lc-h1743.2011.04}. The first outburst begun in October
2008 (3 in the table) has clear signatures of an S outburst,
but, at the same time, a low intensity in the soft 1.3--3 keV
energy band (in this band the outburst was observed by
RXTE/ASM). As can be seen from Fig.\,\ref{fig:lc-h1743.2008.10},
the data point that gave a flux of $160\pm34$ mCrab in the table
is most likely a random outlier. The softness is near the lower
boundary of the range of its variations for the S outbursts. The
second outburst (8, with its peak in April 2011) also formally
resembles an S outburst, but in the soft bands, were it not for
the short (3--4 days) radiation pulse, it would be much closer
to the H outbursts.

Finally, the outburst designated as M (micro-outburst), number 7
in the table, begun in January 2011 was distinguished both by an
anomalously low amplitude in the soft bands ($39\pm8$ mCrab),
which could still be attributed to its hardness, and by a very
low flux in the hard band, $\la42$ mCrab ($3\sigma$ upper
limit), which does not allow us to assign it to the hard H
outbursts. It should be noted, however, that the SWIFT/BAT data
during this outburst were of a rather low quality.\\ [-1mm]

\subsection*{Outbursts of GX 339-4}
\begin{figure*}[t]
\centerline{\includegraphics[width=0.96\textwidth]{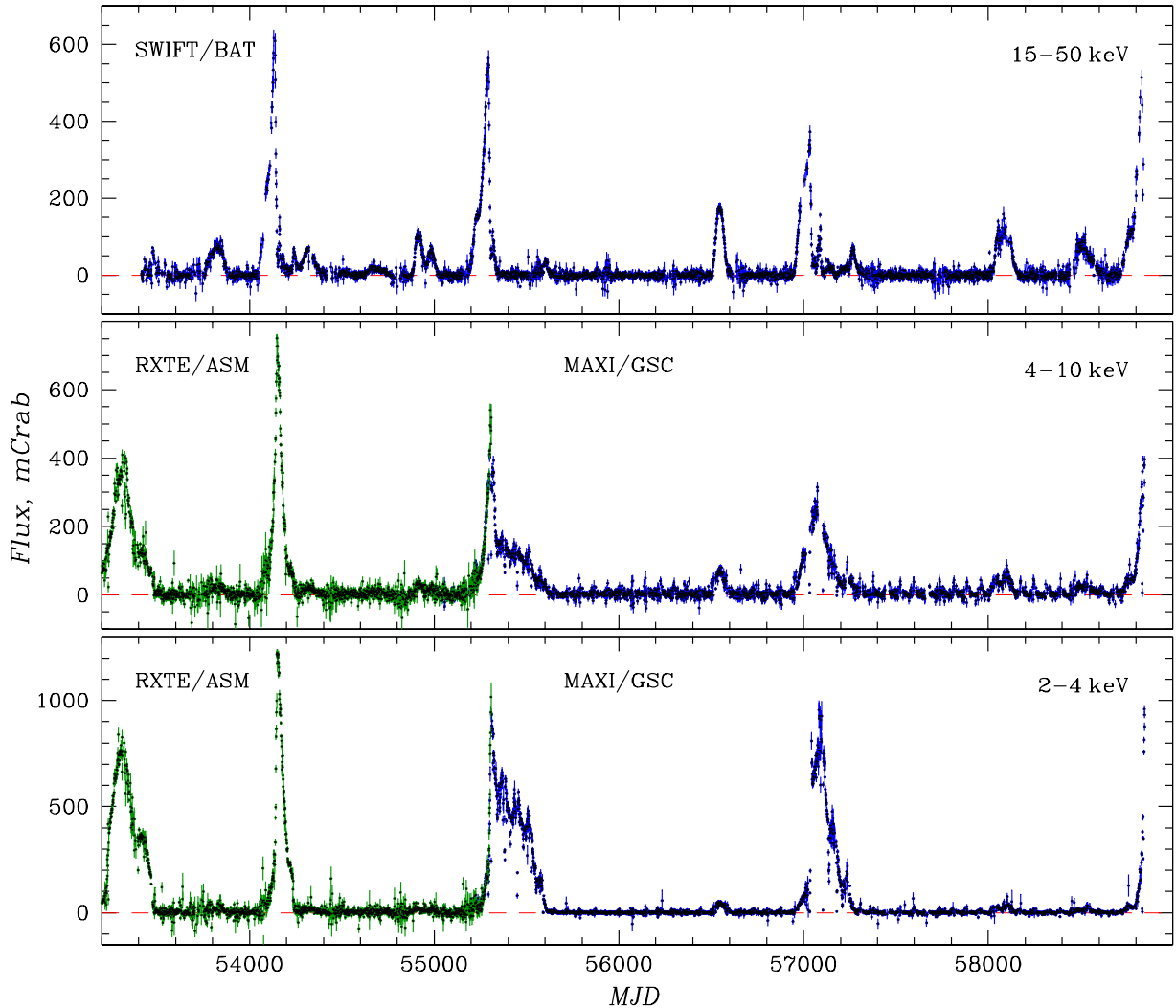}}

\caption{\rm Long-term light curves of the X-ray nova GX\,339-4
  in three energy bands according to the SWIFT/BAT, RXTE/ASM, and MAXI/GSC
  measurements from January 2005 to December 2019.
\label{fig:lc-gx339}}
\end{figure*}

\noindent
Figure\,\ref{fig:lc-gx339}, which is analogous to
Fig.\,\ref{fig:lc-h1743}, shows long-term light curves of the
X-ray transient GX\,339-4.  In 15 years the source experienced
11 outbursts, i.e., they occurred a factor of 1.5 more rarely
than those in the X-ray nova H\,1743-322. Below we will not
consider the earliest outburst observed in MJD\,53200--53500,
because it was not recorded in the hard band (the SWIFT
satellite had not yet begun to operate in its orbit). The
characteristics of the remaining ten outbursts are given in
Table\,2.

The table and Fig.\,\ref{fig:lc-gx339} show that the outbursts of
\mbox{GX\,339-4}, like the outbursts of the X-ray nova H\,1743-322,
were distinguished by a great variety. In particular, both soft
(S or U) and hard (H) outbursts were also present among
them. Figure\,\ref{fig:lc-gx339.2010} and farther
Fig.\,\ref{fig:lc-gx339.2013} below give examples of light
curves for such outbursts with a much better time
resolution. The time profiles of the outbursts that reached the
maximum of their hard radiation in April 2010 (5 in the table,
type S) and September 2013 (6 in the table, type H) are shown.

The outburst in Fig.\,\ref{fig:lc-gx339.2010}, just as all other
soft outbursts, was several times brighter than the similar
outbursts of H\,1743-322, slightly softer ($SHR
\simeq1.74$--1.87) and noticeably longer than them (300--500
days). It follows from Table 2 (columns 7 and 9) that the peak
flux in the soft band during these outbursts was
reached 15--20 (and for outburst 7 almost 50) days later than in
the hard one. Therefore, using $SHR$ for their description looks
somewhat formal.  Nevertheless, this parameter characterizes the
softness of the X-ray radiation from the outbursts quite well
and we will continue to use it to maintain the commonality of
their analysis in these two sources.  Outburst 7 was attributed
to the ultrasoft (U) outbursts, because its $SHR$ reached
$2.56\pm0.16$. On the whole, however, the S and U outbursts of
\mbox{GX\,339-4} have much in common with the giant outburst
H\,1743-322 in 2003. In particular, they exhibit pronounced
phases of initial hard (phase~I), extended soft (phase~II), and
final hard (phase~III) spectral states of the source.

\begin{figure*}[p]
\centerline{\includegraphics[width=0.98\textwidth]{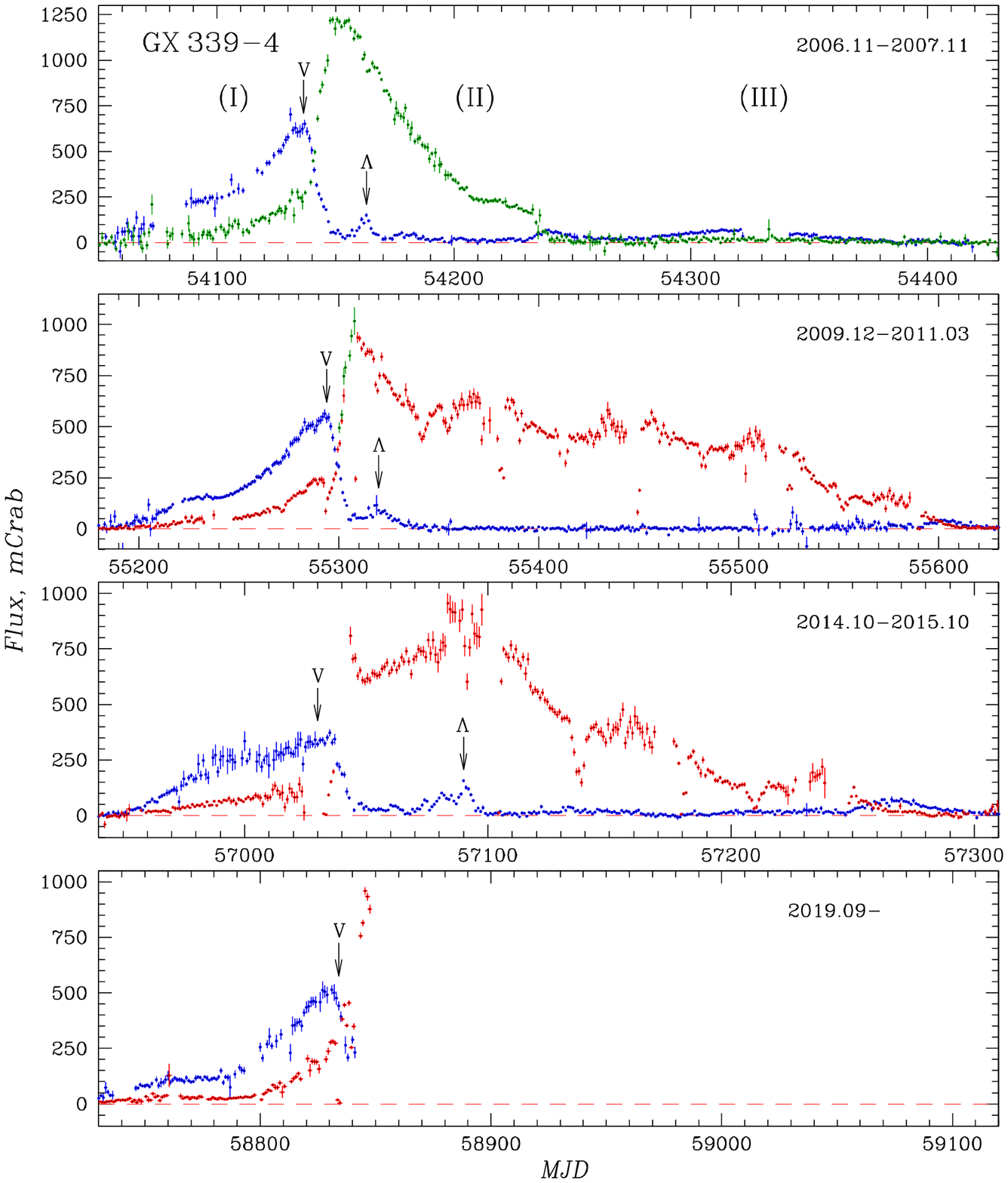}}

\caption{\rm Light curves for the S and U outbursts of the X-ray
  transient GX\,339-4 according to the measurements in the hard
  (15--50 keV, by SWIFT/BAT --- blue points) and soft (1.3--3 or
  4--12 keV, by RXTE/ASM --- green points and MAXI/GSC --- red
  points) bands.\label{fig:lc-gx339sfl}}
\end{figure*}

Figure\,\ref{fig:lc-gx339sfl} compares the light curves for all
soft (S and U) outbursts of GX\,339-4 in two energy bands: hard
15--50 keV (blue circles) and soft 2--4 keV (red or green
circles). The Roman numerals on the upper panel indicate the
phases of the light curve. Interestingly, during phase I (hard
state) the light curves of the source in the soft and hard bands
rise in a correlated way, identically. This means that (1) some
fraction of the soft X-ray radiation is emitted by the source
even in this state and (2) it most likely originates in the same
region where the hard radiation does. It can also be seen that
phase II of the soft state (the outburst in the soft band)
differs greatly for different outbursts in duration, time of
reaching the peak flux, and light curve shape itself. During this phase
the hard flux with a delay of 20--30 days relative to the onset
of the soft outburst drops almost to zero.

In another 10--20 days, during phase II (near the peak of the
soft flux), a short-term manifestation of activity is observed
on the light curves of these outbursts in the hard band --- a
faint burst of hard radiation. Such bursts are indicated in the
figure by the arrows denoted by $\Lambda$. This activity may be
associated with a rise in the temperature of the inner blackbody
disk regions (Shakura and Sunyaev 1973) or with a manifestation
of the hot corona above the cold blackbody disk (Galeev et
al. 1979). Thus, it can have an origin different from the origin
of the hard Comptonized radiation of the source observed during
phases I and III.

Note also the dip at the boundary of phases I and II in the soft
light curve with a duration $\sim10$ days present in all soft
outbursts. The dips are indicated in Fig.\,\ref{fig:lc-gx339sfl}
by the arrows denoted by V. In view of the obvious similarity
between the light curves in the hard and soft bands during phase
I, it can be assumed that the radiation in the soft band has a
different nature at different outburst phases. During phase I it
is closely related to the hard spectral component --- is its
extension (for example, a power-law one) and, accordingly, is
formed in the same high-temperature region; it disappears
simultaneously with the hard component. During phase II it has a
completely different nature and is formed in the geometrically
separated cold opaque region. The dip stems from the fact that
when the soft radiation correlated with the hard one has already
disappeared, the soft blackbody flux has not yet reached the
value required for a smooth monotonic behavior of the light
curve. To some extent its appearance is a result of the choice
of a fairly narrow soft band. The bolometric light curve should
most likely be smooth, while the missing radiation during the
dip is emitted in the intermediate 4--15 keV band.
\begin{table*}[t] 
\caption{Parameters of the outbursts of the X-ray nova GX\,339-4
  from the SWIFT/BAT, RXTE/ASM, and MAXI/GSC data\label{table:gx339}}

\vspace{6mm}
\small
\begin{tabular}{c|r@{--}l|c|c|c|c|r@{}l|c|r@{}l|c|c}\hline
\multicolumn{3}{c|}{}& \multicolumn{2}{c|}{}&  &  &  &  & & &  & & \\ [-3mm]

\multicolumn{3}{c|}{Outburst\aa}&\multicolumn{2}{c|}{$T_{\rm m}(H)$\bb}&\multicolumn{1}{c|}{$\Delta  T$\cc}&$T_{\rm d}$\dd&\multicolumn{2}{c|}{$F_{\rm m}(H)$\ee}&\multicolumn{1}{c|}{$T_{\rm m}(S)$\ff}&\multicolumn{2}{c|}{$F_{\rm m}(S)$\ee}&$SHR$\gg&Type\hh\\ \cline{1-12}

&\multicolumn{2}{c|}{}& & & & & \multicolumn{2}{c|}{}& &\multicolumn{2}{c|}{}& & \\ [-3mm]

N&\multicolumn{2}{c|}{MJD}&\multicolumn{1}{c|}{YYYY--MM}&MJD&days&days&\multicolumn{2}{c|}{mCrab}&days&\multicolumn{2}{c|}{mCrab}&&\\ \hline

&\multicolumn{2}{c|}{}& & & & &  \multicolumn{2}{c|}{}& & \multicolumn{2}{c|}{}& &\\ [-3.5mm]

1& \multicolumn{2}{c|}{2}&\multicolumn{1}{c|}{3}&\multicolumn{1}{c|}{4}& \multicolumn{1}{c|}{5}&6 & \multicolumn{2}{c|}{7}&8& \multicolumn{2}{c|}{9}& 10&11\\  \hline 

&\multicolumn{2}{c|}{}& & & & &  \multicolumn{2}{c|}{}& &  \multicolumn{2}{c|}{}& &\\ [-2.6mm]
  1&53759&53878&2006--04&53846& 119& --   & $ 89.98$&$\pm  8.31$  &--        &$      <81$&$(3\sigma)$&$<0.90\,(3\,\sigma)$&H\\
  2&54060&54385&2007--02&54131& 325& 301& $702.26$&$\pm38.39$ &54150&$1223.25$&$\pm17.14$&$1.74\pm0.10$&S\\
  3&54631&54758&2008--07&54675& 127& 571& $  35.64$&$\pm  6.33$ &--        &$      <33$&$(3\sigma)$&$<0.93\,(3\,\sigma)$&H\\
  4&54884&55023&2009--03&54912& 139& 253& $116.64$&$\pm11.52$ &--        &$    47.53$&$\pm13.26$&$0.41\pm0.12$&H\\
  5&55178&55632&2010--04&55293& 454& 294& $563.80$&$\pm20.25$ &55308&$1016.73$&$\pm67.96$&$1.80\pm0.14$&S\\
  6&56508&56604&2013--09&56543&   96&1330&$178.40$&$\pm8.10$ &--         &$     53.39$&$\pm 5.70$ &$0.30\pm0.03$ &H\\
  7&56946&57300&2015--01&57035& 354& 438& $372.45$&$\pm16.07$&57083&$   955.20$&$\pm45.15$&$2.56\pm0.16$&U\\
  8&58020&58160&2017--11&58082& 140&1074&$158.51$&$\pm20.53$&58104&$     76.18$&$\pm18.06$&$0.48\pm0.13$ &H\\
  9&58475&58599&2018--12&58483& 124&  455&$108.65$&$\pm34.09$&58537&$     34.57$&$\pm10.45$&$0.32\pm0.14$ &H\\
10{\footnotesize $^*$}&58716&\underline{58841}&2019--12&58831&\underline{125}&241&$513.85$&$\pm21.03$&58846&$959.48$&$\pm18.09$ &$1.87\pm0.08$&S\\
\hline
\multicolumn{14}{l}{}\\ [-1mm]
\multicolumn{14}{l}{\aa\ \footnotesize The outburst number and start-end dates (MJD).}\\
\multicolumn{14}{l}{\bb\ \footnotesize The year, month, and date (MJD) of
  reaching the peak flux in the outburst in the H band.}\\ 
\multicolumn{14}{l}{\cc\ \footnotesize The outburst duration.}\\
\multicolumn{14}{l}{\dd\ \footnotesize The recurrence time (the interval between the onset of this outburst and the onset of the previous one).}\\ 
\multicolumn{14}{l}{\ee\ \footnotesize The peak flux in the hard (H: 15--50 keV) and soft (S: 2--4 or 1.3--3 keV during the MAXI/GSC and}\\
\multicolumn{14}{l}{\ \ \   RXTE/ASM observations, respectively) bands.}\\
\multicolumn{14}{l}{\ff\ \footnotesize The date (MJD) of reaching the peak
  flux in the outburst in the S band.}\\ 
\multicolumn{14}{l}{\gg\ \footnotesize The radiation softness (the ratio of the fluxes in the S and H bands).}\\
\multicolumn{14}{l}{\hh\ \footnotesize The type of outburst (hard -- H, soft --  S, ultrasoft -- U).}\\
\multicolumn{14}{l}{\footnotesize $^{*}$\  The underlined values are a lower limit, because the outburst has not finished.}\\ 
\end{tabular}
\end{table*}

\begin{figure*}[t]
\centerline{\includegraphics[width=0.86\textwidth]{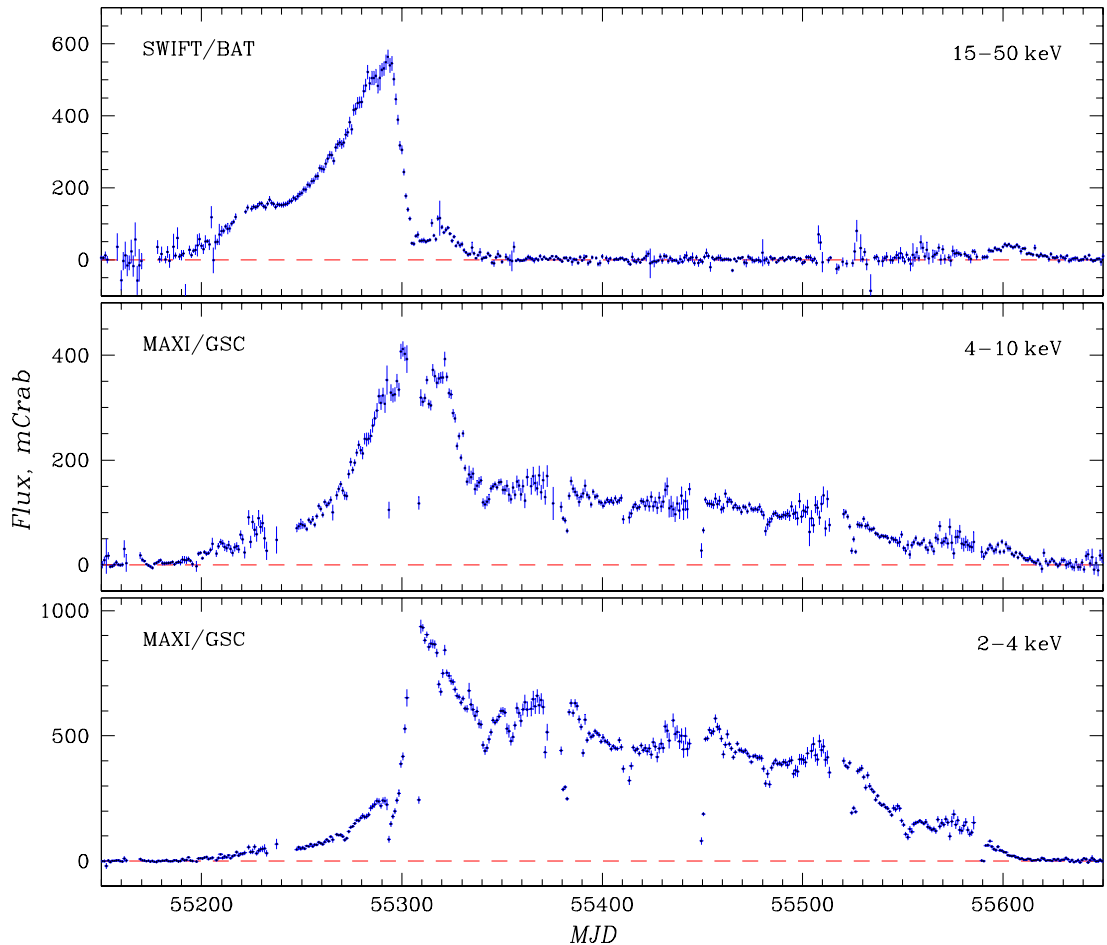}}

\caption{\rm Light curves for the outburst of the X-ray
  transient GX\,339-4 in 2010 (outburst 4 in Table 2, type S) in
  various energy bands (according to the SWIFT/BAT and MAXI/GSC
  measurements).
\label{fig:lc-gx339.2010}}
\end{figure*}

The light curves of the outburst presented in
Fig.\,\ref{fig:lc-gx339.2013}, along with the light curves for
other hard outbursts of this source, are very similar to those
for the hard outbursts of the X-ray nova H\,1743-322. However,
their profile has a more symmetric (triangular) shape with
approximately the same rise and decay phases. The peak flux of
the hard outbursts changed from outburst to outburst by a factor
of 2--3, while the softness of the radiation from the source
($SHR\simeq0.41-0.48$, column 10 in Table\,2) during some
outbursts exceeded that for the hard outbursts of H\,1743-322
($SHR\simeq0.22-0.35$) by a factor of $\sim1.5$. The duration of
the outbursts also exceeded that of the H outbursts in
H\,1743-322, on average, by a factor of 1.5--2.

Curiously, the hard outbursts of this source always precede the
onset of a soft outburst, which differs greatly from the
behavior of the outbursts in the X-ray nova H\,1743-322. In this
case, whereas the delay between the hard and soft outbursts at
the beginning was $\sim200-300$ days, in the succeeding
outbursts it already reached $\sim500-700$ days, while two hard
precursor outbursts at once were observed in the last outburst,
which just reached the peak flux at the end of our
observations. Clearly, this correlation of the H and S outbursts
is by no means a chance one and there must be some hidden
connection between them.  However, we have no reason to directly
attribute them to one outburst event, as was done with the faint
bursts of hard radiation observed on the descent of the light
curves for the powerful S events (2, 5, and 7 in the table and
Fig.\,\ref{fig:lc-gx339sfl}).

\begin{figure}[t]
\centerline{\includegraphics[width=0.99\linewidth]{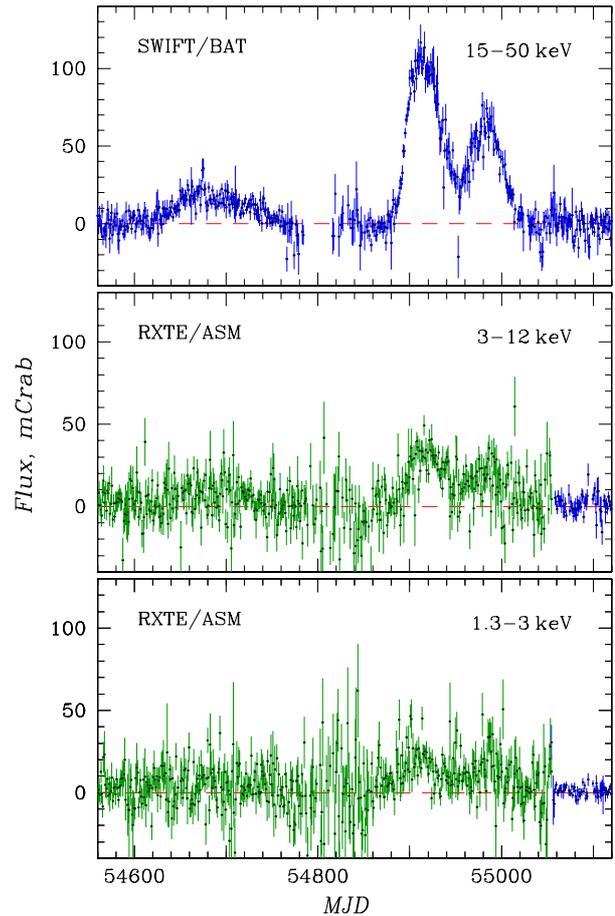}}

\caption{\rm Light curves for the composite outburst of the
  X-ray transient GX\,339-4 in 2009 (outburst 4 in Table\,2,
  type H) in various energy bands (according to the SWIFT/BAT
  and RXTE/ASM measurements).  The faintest recorded outburst of
  the source begun in July 2008 (outburst 3, type H) is also
  shown.\label{fig:lc-gx339.2009}}
\end{figure}

Outburst 4 from Table\,2 begun in March 2009 is of
interest. Although it undoubtedly belongs to the H type (it was
hardly recorded in two soft energy bands), in the hard band (see
Fig.\,\ref{fig:lc-gx339}) its profile has a pronounced
two-component shape that, as we saw above using H\,1743-322 as
an example, is one of the signatures of S outbursts. In
Fig.\,\ref{fig:lc-gx339.2009} the light curves measured during
this outburst are presented with a better time
resolution\footnote{The same figure shows the previous very
  faint hard outburst 3 occurred in July 2008.}. We see that the
two-component profile of outburst 4 actually bears no
resemblance to the profile of the S outbursts in H\,1743-322
(the second component has the appearance of not a step, but a
new, almost equally powerful outburst in all bands). In this
case, we most likely deal with a superposition of two H
outbursts. Below we will show that this is actually the case.

\section*{DISCUSSION}
\noindent
Apart from the formal difference in maximum attained
accretion rate, the observed differences in the
outbursts of X-ray novae can be explained by the
presence (or absence) of a soft component in the
broadband spectrum of their X-ray radiation associated
with the outer cold, optically thick accretion
disk region radiating like a blackbody (Shakura and
Sunyaev 1973).

\subsection*{Accretion Disk Evaporation}
\noindent
The cold opaque region should always be present
in the disk --- at least on its far periphery, where
the sudden infall or accumulation of matter being
transferred from the normal star occurs (in this case,
however, the radiation from the cold region falls into
the optical and infrared range). We are interested in
the X-ray radiation and, therefore, the question is how
closely toward the black hole this region extends, i.e.,
at what radius the transition from the cold disk to the
high-temperature inner region occurs.

Evaporation of the cold disk (Meyer et al. 2000) under
conditions when the mechanisms of heat removal in this disk (by
radiation or heat conduction) cannot cope with the viscous
release of the gravitational energy of accreting matter growing
toward the disk center can serve as a cause of this
transition. Since at the first stage of outburst development
(Lyubarskii and Shakura 1987) from the matter accumulated on the
periphery not as much of it flows toward the black hole --- an
initial disk (``tongue'') of relatively low surface density is
formed, the formation of an extended high-temperature region,
radiating in the hard X-ray band near the black hole should not
cause a surprise. Subsequently, the accretion rate grows, the
surface density of the matter in the disk and its optical depth
increase. This allows the heat to be removed more efficiently by
radiation; accordingly, the area of the cold disk region grows,
while that of the hot one decreases.

The closer the cold disk approaches the black hole, the greater
the fraction of the gravitational energy of its matter released
during accretion is irradiated in the soft blackbody component
of the broadband spectrum of the X-ray nova, and the higher the
temperature of this radiation must be. Accordingly, the fraction
of the energy radiated in the inner high-temperature zone must
decrease.

\subsection*{Position of the Truncation Radius of the Cold Disk}
\noindent
A verification of the fact that the ratio of energy release in
the two parts of the disk changes could be a confirmation of
this picture of outburst development in the truncated disk
model. After all, an alternative variant of outburst
development, where the radius separating the accretion disk into
two components remains constant, while the observed variability
is related exclusively to the change in accretion rate and the
corresponding change in cold disk temperature, is possible in
principle. The temperature of the inner disk zone in this case
can also change with accretion rate, but not the fraction of the
energy being release in it. Besides, completely different
origins of the hard X-ray radiation during accretion onto a
black hole are discussed --- its formation in the hot, optically
thin corona forming above the cold disk surface (Galeev et
al. 1979; Haardt and Maraschi 1991) or in the relativistic jets
emitted from the core of a black hole and observed in a number
of sources in the radio, optical, and soft X-ray bands (see,
e.g., Espinasse and Fender 2018). 

The energy redistribution in the radiation spectra of X-ray
novae (a change in the ratio of energy release in the two parts
of the disk) is clearly seen from the shape of the light curves
for their outbursts in the hard band: in comparison with the
regular (FRED) light curve of H outbursts, the light curves of S
and U outbursts are distorted starting from some time --- a dip
is formed in them, as if a certain fraction of their luminosity
suddenly disappears, --- and are restored at a level expected
for the regular light curve only much later. It is clear where
this fraction of the luminosity in the hard band vanishes --- a
soft X-ray outburst of the source begins exactly at this time.

\subsection*{Bolometric Light Curves}
\begin{figure*}[t]
\begin{minipage}{0.55\textwidth}
  \includegraphics[width=0.98\textwidth]{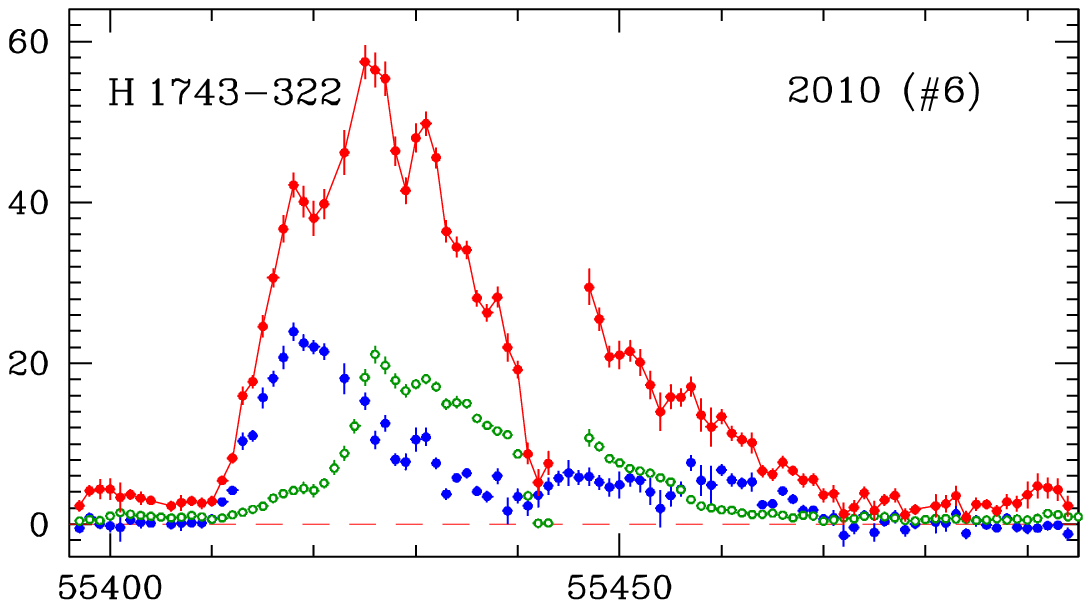}
\end{minipage}
\hspace{-4mm}\begin{minipage}{0.55\textwidth}
  \includegraphics[width=0.90\textwidth]{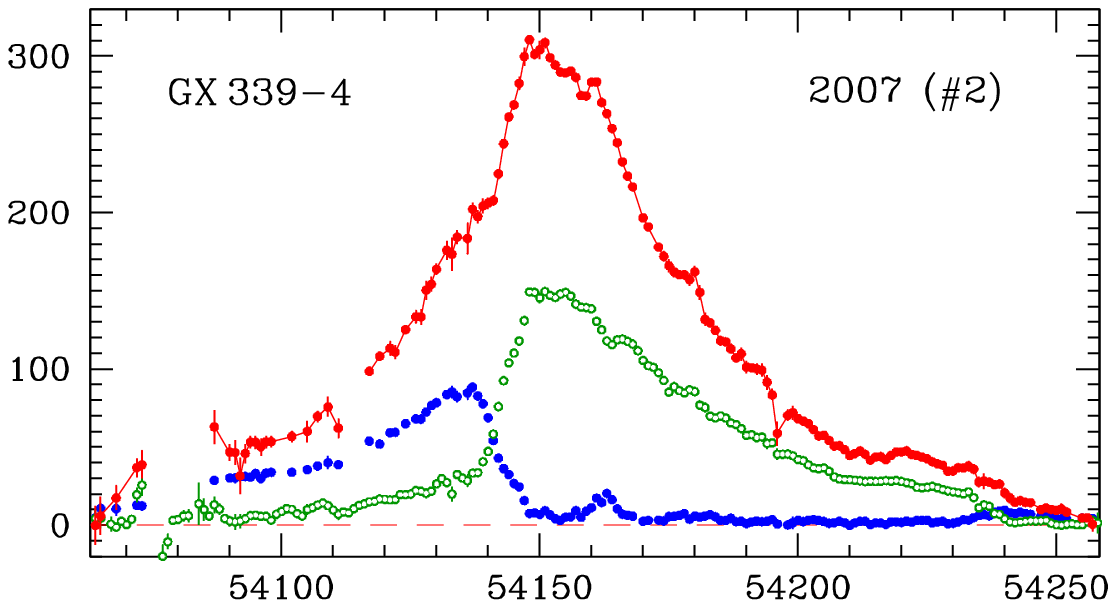}
\end{minipage}\\
\begin{minipage}{0.55\textwidth}
  \includegraphics[width=0.98\textwidth]{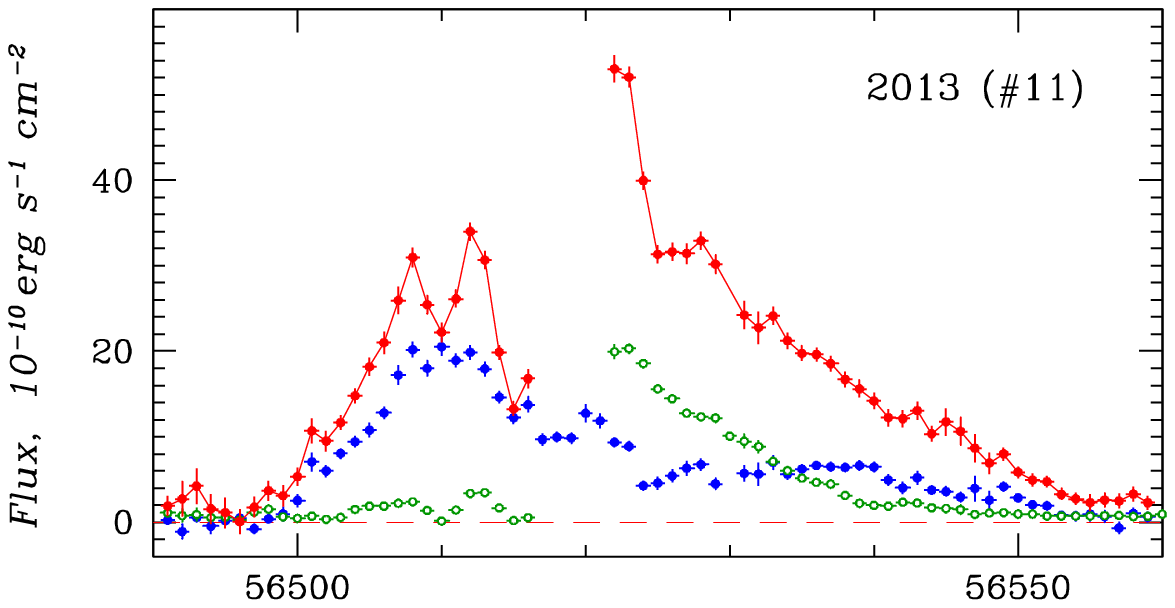}
\end{minipage}
\hspace{-4mm}\begin{minipage}{0.55\textwidth}
  \includegraphics[width=0.90\textwidth]{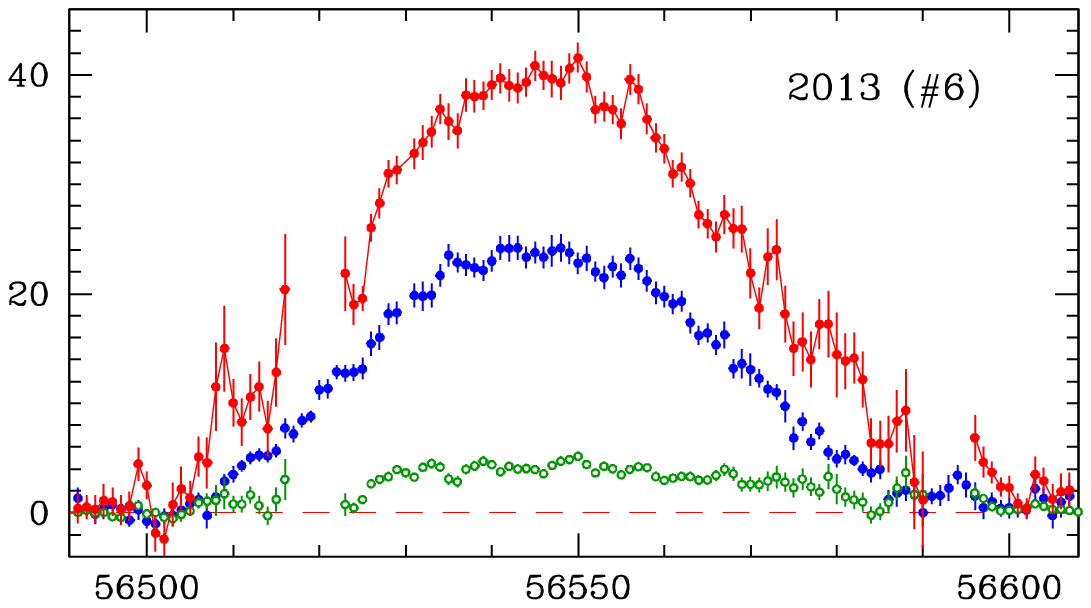}
\end{minipage}\\
\begin{minipage}{0.55\textwidth}
  \includegraphics[width=0.98\textwidth]{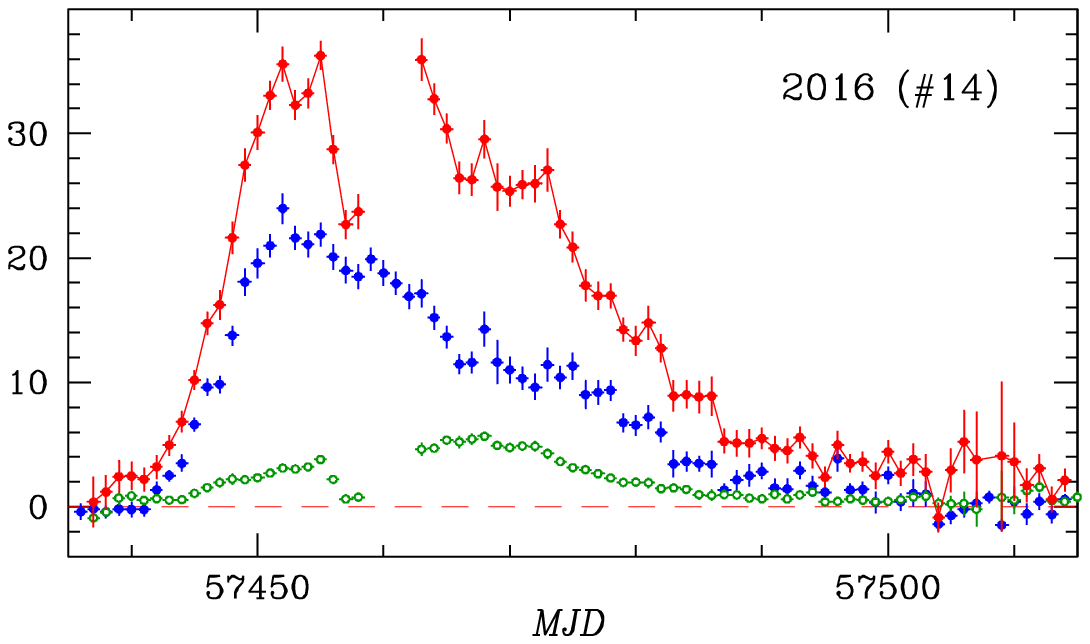}
\end{minipage} 
\hspace{-4mm}\begin{minipage}{0.55\textwidth}
  \includegraphics[width=0.90\textwidth]{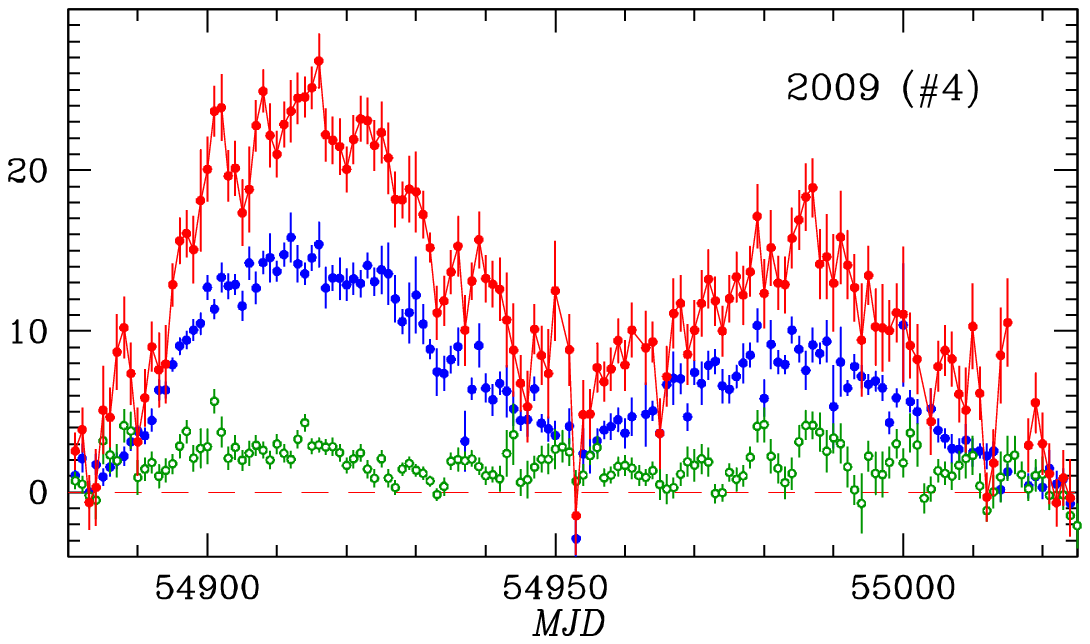}
\end{minipage}
\vspace{-4mm}

\caption{\rm Broadband (2--50 keV, the red circles and the solid
  line) light curves for the outbursts of the X-ray novae
  H\,1743-322 (left) and GX\,339-4 (right) obtained by adding
  the 2--4, 4--15, and 15--50 keV fluxes (converted from the
  photon fluxes via the spectrum of the Crab nebula). The 15--50
  keV (blue filled circles) and 2--4 keV (green open circles)
  fluxes based on the SWIFT/BAT, RXTE/ASM, and MAXI/GSC
  measurements are also shown.\label{fig:cr_bolom}}

\vspace{-1mm}
\end{figure*}

\noindent
It is interesting how smoothly the accretion rate onto the black
hole changes during such a transition --- how the bolometric
luminosity of the source behaves.  In Fig.\,\ref{fig:cr_bolom}
we attempted to answer this question by reconstructing the
broadband X-ray light curves of the X-ray novae H\,1743-322
(left) and GX\,339-4 (right) during some of their S and H
outbursts. For this purpose, the photon fluxes (in mCrab) from
these novae in the three 2--4, 4--10, and 15--50 keV energy
bands\footnote{The outbursts of GX 339-4 in 2007 and 2009 were
  investigated with RXTE/ASM; accordingly, the soft 1.3--3 and
  3--12.2 keV bands were used in constructing the light curves
  of these outbursts.} were converted to the radiation fluxes
(in $\mbox{erg s}^{-1}\ \mbox{cm}^{-2}$) via the spectrum of the
Crab nebula.  The photon spectrum of the nebula was specified in
the form
$$F(E) = 9.8\times (E/1\ \mbox{keV})^{-2.1}\ \mbox{phot
  s}^{-1}\ \mbox{cm}^{-2}\ \mbox{keV}^{-1}.$$ The 4--10 keV flux
was converted to the 4--15 keV flux. As a result, an estimate of
the flux from the X-ray novae in a wide continuous range, 2--50
keV, was obtained.

We see that the bolometric light curves for the S (in 2010 and
2013) and H (in 2016) outbursts of H\,1743-322 reconstructed in
this way were in many respects similar --- had the same profile
shape (FRED) and differed only by the normalization (the H
outburst was fainter). The transition from the hard state to the
soft one in the S outbursts was not accompanied by any features
in the bolometric light curves. The same can also be said about
the bolometric light curves of the X-ray transient GX 339-4 ---
the light curve of its soft outburst in 2007 has a smooth dome
shape, nothing in it points to the transition from the hard
state to the soft one. The light curves of the hard outbursts
occurred in 2009 and 2013 in many respects resemble this light
curve, but are characterized by a lower brightness and a flatter
top.
\begin{figure}[t]
\centerline{\includegraphics[width=0.99\linewidth]{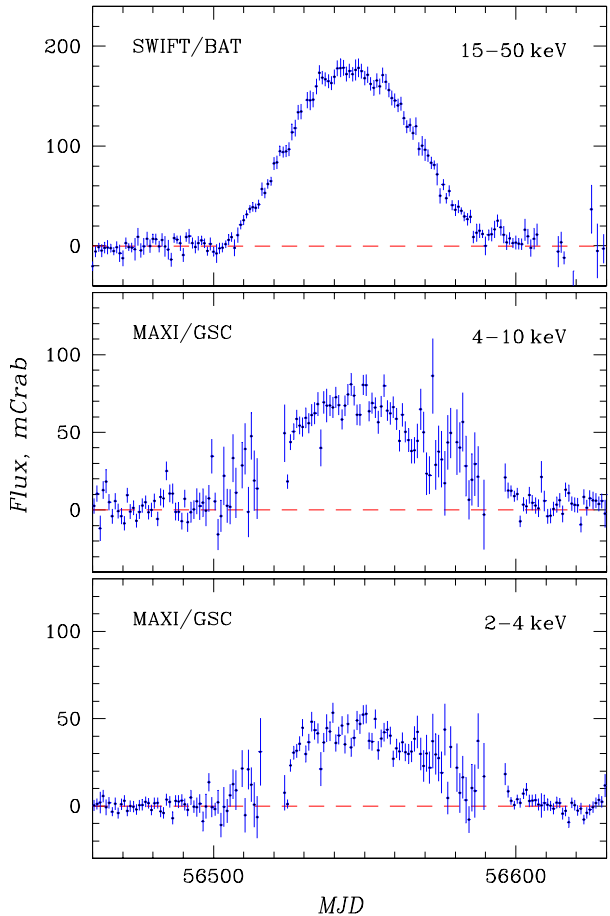}}

\caption{\rm Light curves for the outburst of the X-ray
  transient GX\,339-4 in 2013 (outburst 6 in Table\,2, type H) in
  various energy bands (according to the SWIFT/BAT and MAXI/GSC
  measurements).\label{fig:lc-gx339.2013}}
\end{figure}

The presented light curves of GX\,339-4 slightly differ in shape
from the light curves of H\,1743-322: their rise and decay
phases have almost the same slope and duration\footnote{This
  definitely does not pertain to the 2009--2011 outburst of this
  source, which had a highly asymmetric shape, but is apparently
  true for the 2014--2015 outburst as well (see
  Fig.\,\ref{fig:lc-gx339sfl}).}. According to the self-similar
solution of the nonlinear partial differential equation that
describes unsteady disk accretion onto a black hole (Lyubarskii
and Shakura 1987), this is possible if the scattering by
electrons dominates in the opacity of matter in the region of
main energy release at both phases. In this case, the accretion
rate changes with time as $\dot{M}\sim t^{\alpha},$ where the
exponent $\alpha_{\rm r} \simeq 1.67$ at the rise phase and
$\alpha_{\rm d}\simeq -1.63$ at the decay phase, when the
remnants of the fallen matter flow down over the disk. If the
bremsstrahlung processes dominate in the opacity of matter, then
$\alpha_{\rm r}\simeq 2.47$ and $\alpha_{\rm d}\simeq -1.25$
(Lyubarskii and Shakura 1987). Given the power of the outbursts
in GX\,339-4, the dominance of scattering in the opacity seems
natural.

\subsection*{Features in the Light Curves}
\noindent
As has already been noted, the transition between the states in
the soft outbursts of both sources did not lead to clear
features in their bolometric light curves. At the same time,
there were features (``dips'' and ``knees'' of various
intensities) in the light curves at other times (see
Fig.\,\ref{fig:cr_bolom}). They are also seen in the light
curves in the narrow 2--4 and 15--50 keV bands obtained by
different instruments, i.e., they are definitely not an
instrumental effect. A strong dip shortly after the passage of
the peak flux, along with several fainter features, can also be
seen in the light curve for the hard outburst of H\,1743-322
occurred in 2016. Dips and bursts are also seen in the light
curves for the hard outbursts of the X-ray transient
GX\,339-4. In the light curve for the soft outburst of the
source in 2007 these features are less pronounced, possibly, due
to its high brightness.

These features can be formed due to the accretion flow
instabilities arising in the disk, primarily in its inner
radiation-dominated part (Shakura and Sunyaev 1976). Although
the disk averaged accretion rate remains constant, it can
undergo significant changes locally and on a short time scale
due to the instability.

Another reason for the appearance of ``steps'' and dips in the
bolometric light curve can be the transition in the opacity of
disk matter from the Thomson regime, whereby the scattering by
electrons dominates in the matter-radiation interaction, to the
bremsstrahlung one, when the free-free absorption processes
begin to lead (Lipunova and Shakura 2000).

The largest feature (a dip or a step) with a duration $\sim10$
days observed in the light curves of all three outbursts may
also be related to some large-scale nonuniformity of the flow of
matter in the disk, for example, the passage of a ring
perturbation --- a wave containing much residual matter, and an
eclipse of the region of main energy release (the inner regions
of the cold disk and partly the hot cloud) by it.

\subsection*{2009 Outburst}
\noindent
The 2009 outburst of GX\,339-4 has a two-component light curve
that resembles the light curve of an S outburst in the hard
band. Previously we have already noted that a superposition of
two H outbursts rather than one S outburst is most likely
observed in this case. The bolometric light curve of this event
in Fig.\,\ref{fig:cr_bolom} confirms this assumption: the
two-component light curve of this outburst was retained and even
enhanced, while the corresponding dips in the hard light curves
of the S outbursts are smeared and virtually disappear in their
bolometric light curves (this is particularly clearly seen in
the light curves for the outbursts of H\,1743-322 occurred in
2010 and 2013).

\subsection*{Soft Radiation in the Hard State}
\noindent
We have already noted above that during phase I of
S-outburst development in GX\,339-4 its light curves
in the soft and hard bands behaved in a correlated way.
This can also be clearly seen from Fig.\,\ref{fig:cr_bolom} (the 2007
outburst). Such a behavior can be explained only by
assuming that the observed soft X-ray radiation from
the source is formed in the same disk region as its
hard radiation, i.e., in the inner hot zone. Figure\,\ref{fig:cr_bolom}
shows that during their hard H outbursts, when the
boundary of the cold accretion disk region recedes far
from the black hole, both X-ray novae GX\,339-4 and
H\,1743-322 also emit a noticeable soft X-ray flux.
Obviously, this radiation must be formed in the high-temperature
cloud surrounding the black hole.

Grebenev (2020) showed that the X-ray spectrum forming in a
cloud of high-temperature plasma, when its intrinsic
low-frequency radiation is Comptonized, must behave precisely in
this way. As a result of detailed numerical computations
including both Compton photon scattering and bremsstrahlung
photon production and absorption processes in the plasma, he
showed that the hard power-law spectrum typical for X-ray novae
naturally extends to the infrared, optical, and ultraviolet
ranges and only at lower energies does it transform into a
rapidly decaying Rayleigh-Jeans spectrum. Therefore, the soft
X-ray radiation observed during the hard state of the nova must
be simply a power-law extension of the hard X-ray radiation from
the source.

\subsection*{Nature of the Two Types of Outbursts}
\noindent
We saw that the spectral state of the source at a specific time
is largely determined by the position of the cutoff radius (the
radius at which the cold opaque disk evaporates and transforms
into a high-temperature semitransparent disk). However, it is
unclear whether this position depends only on the accretion rate
onto the black hole or also on the entire history of the X-ray
outburst development. It is also unclear what dictates the
choice of the type of outburst realized in a given accretion
episode.

\begin{figure*}[t]
\centerline{\includegraphics[width=0.8\textwidth]{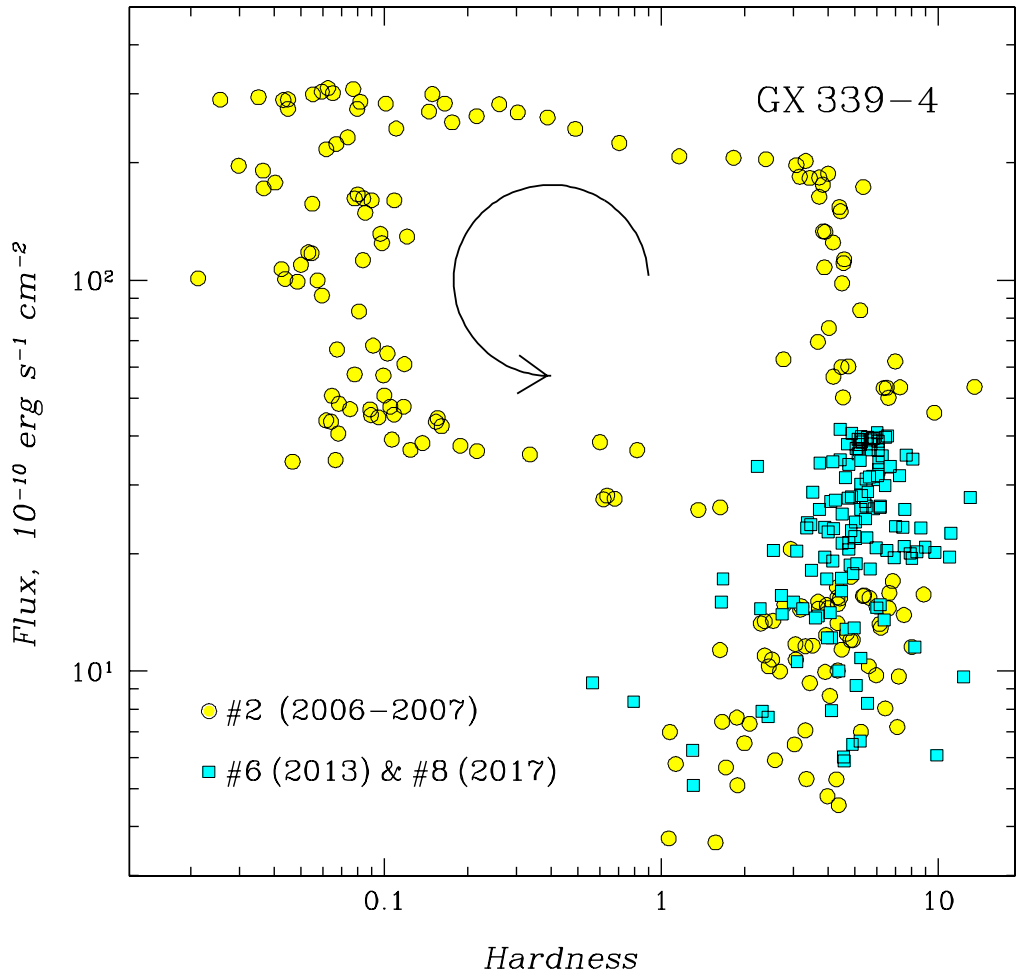}}

\caption{\rm Comparison of the hardness-intensity diagrams for a
  soft (type S, yellow circles) and two hard (type H, blue
  squares) outbursts of the X-ray transient GX\,339-4. The soft
  outburst was observed by SWIFT/BAT and RXTE/ASM in 2006--2007
  (2 in Table 2), the hard ones were observed by SWIFT/BAT and
  MAXI/GSC in 2013 and 2017 (6 and 8 in Table 2). The hardness
  was defined as the ratio of the 15--50 and 2--4 keV fluxes,
  the intensity is the integrated 2--50 keV flux (the values of
  the light curves in Fig.\,\ref{fig:cr_bolom}). \label{fig:hid-gx339}}
\end{figure*}

Figure\,\ref{fig:hid-gx339} presents hardness-intensity diagrams
for the outbursts of different types observed in the X-ray nova
GX\,339-4 in 2006--2007 (S type, yellow circles) and in 2013 and
2017 (H type, blue squares). We took the ratio $HSR$ of the
15--50 and 2--4 keV fluxes (the measured RXTE/ASM 1.3--3 keV
flux was reduced to the 2--4 keV band) as the hardness and the
integrated 2--50 keV flux as the intensity. The fluxes were
converted from the photon fluxes in the same way as was
described above when constructing Fig.\,\ref{fig:cr_bolom}.
Such diagrams were constructed and studied previously by Belloni
(2010).

The circular arrow in the figure indicates the course of
development of the source's soft outburst.  We see that at the
onset of the outburst the source was in the hard state with a
hardness $HSR\ga1$. At some time, on reaching the peak flux in
the 15--50 keV band, the edge of the cold disk begins to
tangibly shift toward the black hole, leading to a dominance of
the soft blackbody component in the source's X-ray spectrum. The
radiation hardness drops to $HSR\la 0.2$, i.e., the source
passes to the soft spectral state. It is even possible that the
cold edge of the disk reaches the radius of the innermost stable
circular orbit near the black hole ($3\,R_{\rm g}$) and the
high-temperature region disappears completely.  During the
transition between the states the total luminosity of the source
(accretion rate) continues to slowly increase. That is why the
peak flux in the soft band in such outbursts, as a rule, is
reached later and is greater in absolute value that the peak
flux in the hard band (see Fig.\,\ref{fig:lc-gx339sfl}).

After some time the accretion rate begins to drop, the central
zone of the cold disk begins to evaporate, and a cloud of
high-temperature plasma is again formed around the black hole
and begins to increase in size. The spectral hardness rapidly
rises to its original value $HSR\ga1$. It is important that the
reverse transition from the source's soft state to the hard one
occurs at a noticeably (by an order of magnitude) lower
bolometric luminosity (lower accretion rate) than does the
direct transition. Clearly, some time delay dependent on the
history (hysteresis) is present in the outburst
development. However, at the end of its development the outburst
reaches the same curve of the hardness-intensity dependency from
which it started.
 
The behavior of the hard (H) outbursts at the initial stage
differs little from the behavior of the soft ones.  However,
subsequently both hard outbursts considered stopped in their
development and this happened at the very beginning of the loop
formed on this diagram by the soft outburst. After some time
they returned back to the region of low fluxes along the same
hardness curve along which they came.

It seems quite probable that in these accretion episodes too
little matter was transferred to the black hole for a normal
powerful (soft) outburst to be formed. The required accretion
rate, at which a sufficiently extended outer disk with a high
surface density would be formed, was not reached. Since no
hysteresis manifestations were observed, the position of the
cutoff radius of this disk (its edge) either was entirely
controlled by the accretion rate or the disk edge was too far
from the center and the disk radiation had no noticeable effect
on the observed X-ray spectrum of the source.

\section*{CONCLUSIONS}
\noindent
Our analysis of the data from the continuous RXTE, SWIFT, and
MAXI monitoring of the recurrent X-ray novae H\,1743-322 and
\mbox{GX\,339-4} in 2005--2019 shows the following:

\begin{enumerate}
\item Depending on the profile shape of the light curves in the
  soft 2--4 keV and hard 15--50 keV energy bands, the outbursts
  of these X-ray novae can be assigned to one of several
  characteristic types: hard (H), soft (S), ultrasoft (U),
  intermediate (I), and, possibly, micro-outbursts (M).
  
\item The difference between the types of outbursts is
  successfully explained in the truncated accretion disk model
  and is related to the presence (I, S, and U outbursts) or
  absence (H outbursts) of a powerful soft blackbody component
  in the source's broadband X-ray spectrum forming in the outer
  cold opaque accretion disk region.

\item The appearance of a blackbody component in the source's
  X-ray spectrum (phase II of the I, S, and U outbursts) is
  accompanied by a sharp decrease in its hard X-ray flux (thus,
  it is associated not just with a change in the surface
  temperature of the outer disk, but with a shift of its inner
  edge toward the black hole).

\item The soft X-ray component in the spectra of the X-ray novae
  during their hard spectral state (during the H outbursts or
  during phases I and III of the S and U outbursts) is not
  associated with the cold disk region but is formed in the same
  hot inner region as the hard radiation (this is suggested by
  the similarity of the soft and hard light curves for the novae
  in this spectral state noted above (see also Grebenev et
  al. 2016); this component is most likely just an extension of
  the source's hard power-law spectrum, forming in the inner
  disk region as a result of Comptonization, to the softer X-ray
  band).

\item The bolometric light curves for a number of outbursts in
  the X-ray nova GX\,339-4 had a nearly symmetric shape with
  equal rise and decay phases, they differed sharply from the
  light curves of the X-ray nova H\,1743-322 characterized by a
  fast rise and a slow decay; the symmetric shape of the light
  curves can arise from the greater intensity of the outbursts
  in GX\,339-4 and the predominance of Thomson scattering in the
  opacity of accreting matter in its disk.
  
\item The dip in the light curve of the X-ray nova GX\,339-4 in
  the soft 2--4 keV band when passing from phase I to phase II
  of the soft (S and U) outbursts is associated with the change
  in the shape of the nova spectrum when the soft power-law
  X-ray component disappears in it and a soft blackbody X-ray
  component appears (and with the narrowness of the range used
  to construct this light curve).

\item The faint recurrent burst of activity observed in the hard
  15--50 keV light curves for the S and U outbursts of the X-ray
  nova GX\,339-4 during its phase II (simultaneously with the
  attainment of its peak flux in the soft 2--4 keV band) can
  also have a different nature than that of the hard outbursts
  during phases I and III; it can be a consequence of the rise
  in the surface temperature of the cold part of the disk near
  its inner edge (Shakura and Sunyaev 1973) or the appearance of
  a hot corona above the disk (Galeev et al. 1979).

\item The hard outbursts of the X-ray transient GX\,339-4 (one
  or more) always precede its long soft outbursts, whereas the
  outbursts of the X-ray nova H\,1743-322 had a more complex
  distribution --- after the giant 2003 outburst only soft U, S,
  and I outbursts were observed for 5--6 years, and virtually
  only hard outbursts occurred subsequently.
  
\item The choice of which type of outburst is realized at a
  given time seems to depend only on the attained maximum
  accretion rate onto the source and, accordingly, on the total
  mass of the matter transferred during the outburst to the
  black hole. A soft (powerful) outburst develops only in the
  case where the accretion rate reaches some critical value. The
  position of the cutoff (truncation) radius, at which the cold
  opaque disk evaporates and turns into a high-temperature
  semitransparent disk, is determined in these outbursts not
  only by the accretion rate, but also by the entire history of
  the outburst development. In the hard outbursts the position
  of the cutoff radius either is entirely controlled by the
  accretion rate or the cold disk is cut off too far from the
  black hole to have a noticeable effect on the source's
  radiation.
\end{enumerate}
  
\vspace{1mm}

\section*{ACKNOWLEDGMENTS}
\noindent
The study is based on the MAXI data provided by
RIKEN (JAXA) and the MAXI team, the SWIFT data
provided by NASA and the SWIFT team, and the RXTE
data provided by the ASM/RXTE team.

\section*{FUNDING}
\noindent
A.\,G., Yu.\,D., V.\,K., and K.\,O. are grateful to the Space
Research Institute of the Russian Academy of Sciences
for the organization of the work practice during which
this study was performed. S.\,G., I.\,M., and A.P. are grateful to
Basic Research Program no. 12 of the Russian Academy
of Sciences (``Questions of the Origin and Evolution of
the Universe with the Application of Methods of Ground-Based
Observations and Space Research'') and the Russian 
Foundation for Basic Research (project no. 17-02-01079-a) for
their financial support.\\ [4mm]

\small

\vspace{5mm}

\begin{flushright}
{\sl Translated by V. Astakhov\/}
\end{flushright}
\end{document}